\documentclass[useAMS,usenatbib,usegraphicx]{mn2e}
\usepackage{amsmath,amssymb}
\newcommand{\SFR}{\ensuremath{\mathrm{SFR}_{\rm c,ff}}}

\title[Semi-analytic model of the turb. multi-phase ISM]{A semi-analytic model of the turbulent multi-phase interstellar medium}
\author[H. Braun \& W. Schmidt]{H. Braun$^{1}$\thanks{E-mail:
hbraun@astro.physik.uni-goettingen.de} and W. Schmidt$^{1}$\footnotemark[1]\\
$^{1}$Institut f\"ur Astrophysik, Universit\"at G\"ottingen, Friedrich-Hund Platz 1, D-37077 G\"ottingen, Germany}

\begin{document}

\date{Accepted 2011 September 26. Received 2011 August 12; in original form 2011 April 29}

\pagerange{\pageref{firstpage}--\pageref{lastpage}} \pubyear{2011}

\maketitle

\label{firstpage}

\begin{abstract}
We present a semi-analytic model for the interstellar medium that considers local processes and structures of turbulent star-forming gas. A volume element of the interstellar medium is described as a multi-phase system, comprising a cold and a warm gas phase in effective (thermal plus turbulent) pressure equilibrium, and a stellar component. The cooling instability of the warm gas feeds the cold phase, while various heating processes transfer cold gas to the warm phase. The cold phase consists of clumps embedded in diffuse warm gas, where only the molecular fraction of the cold gas may be converted into stars. The fraction of molecular gas is approximately calculated, using a Str\"omgren-like approach, and the efficiency of star formation is determined by the state of the cold gas and by the turbulent velocity dispersion on the clump length scale. Gas can be heated by supernovae and UV-emission of massive stars, according to the evolutionary stages of the stellar populations and the initial mass function. Since turbulence has a critical impact on the shape of the gaseous phases, on the production of molecular hydrogen and on the formation of stars, the consistent treatment of turbulent energy -- the kinetic energy of unresolved motions -- is an important new feature of our model. Besides turbulence production by supernovae and by the cooling instability, we also take into account the forcing by large scale motions.

We formulate a set of ordinary differential equations, which statistically describes star formation and the exchange between the different budgets of mass and energy in a region of the interstellar medium with given mean density, size, metallicity and external turbulence forcing. By exploring the behaviour of the solutions, we find equilibrium states, in which the star formation efficiencies are consistent with observations. Kennicutt-Schmidt-like relations naturally arise from the equilibrium solutions, while conventional star formation models in numerical simulations impose such relations with observed efficiency parameters as phenomenological calibrations. 

Beyond the semi-analytic approach, a potential application is a complete subgrid scale model of the unresolved multi-phase structure, star formation and turbulence in simulations of galaxies or in cosmological simulations. The formulation presented in this article combines various models focusing on particular processes and yet can be adopted to specific applications, depending on the range of resolved length scales.
\end{abstract}


\section{Introduction}
The capabilities of contemporary supercomputing enable us to model the evolution of the baryonic gas in the universe with unprecedented sophistication. Adaptive methods such as smoothed particle hydrodynamics (SPH) and adaptive mesh refinement (AMR) in Eulerian grid codes allow us to cover a huge dynamic range such that simulations of the formation and evolution of galaxies from cosmological initial conditions at high resolution ($\sim 100\,{\rm pc}$) are within reach \citep{GnedKravt10,AgerTey10}. In simulations of isolated disc galaxies, it is feasible to resolve length scales down to $\sim 10\,{\rm pc}$ \citep{AgerLake09,TaskTan09}. Computations on these length scales entail the problem to account for various physical processes in the multi-phase interstellar medium \citep{MayGov08}. Notwithstanding the high numerical resolution that can be achieved, several important processes cannot be fully resolved and have to be described by means of a sub-grid scale (SGS) model. 

The distribution of the gas among the different phases of the ISM is controlled by the following physical processes. The fragmentation of warm neutral gas (number density $n\lesssim 1\,{\rm cm^{-3}}$, temperature $T\gtrsim 10^{4}\,{\rm K}$) is driven by gravitational instabilities on length scales $\sim 0.1\ldots 1\mathrm{kpc}$ \citep[e.g. ][]{Toom64,Wada2002,Kravtsov2003,Li2005,Wada2007,AgerLake09}. The gravitational contraction of gas is supported by cooling processes in converging flows that produce the cold neutral phase ($n\gtrsim 10\,{\rm cm^{-3}}$, $T\lesssim 10^{3}\,{\rm K}$). Gravitational and cooling instabilities, and, possibly, magnetic fields act in concert to form dense star-forming clouds, in which molecular hydrogen is produced at densities $\gtrsim 100\,{\rm cm^{-3}}$ \citep[e.g. ][]{DobbsGlov08,RobertKravt08,TaskTan09}. The radiation from hot massive stars and blast waves from supernovae feeds energy back into the interstellar medium. Gravity, cooling, and stellar feedback are potential drivers of turbulence \citep{ElmeSca04,LowKless04,AviBreit04,BurkGenz09,KlessHenne10,BourElme10,Federrath2011}, which, in turn, has an impact on the stability of the gas \citep{BonaPer92,RomBurk10}. 

In large-scale simulations, where the smallest resolved length scales range from the scale of star-forming regions to galactic scales, it is a major challenge to account for the sub-resolution structure and dynamics of the ISM (recent reviews are given by \citet{McKee2007} and \citet{Hens09}). On the one hand, isolated disc galaxy simulations serve as idealized models of galaxy evolution that avoid some of the difficulties one faces in cosmological simulations \citep{DobbsGlov08,RobertKravt08,AgerLake09,TaskTan09,DobbsPrin10,BourElme10}. Although the highly artificial initial conditions are problematic, isolated discs can be used to study dynamical properties of the ISM at high resolutions and to test advanced models of unresolved processes. Because artefacts may result from discs that are adiabatically unstable, \citet{WangKless10} defined an adiabatic disc that is stable over the rotation time scale. On the other hand, substantial efforts have been made to zoom into halos from cosmological simulations and to re-simulate galaxies evolving from those halos at the highest feasible resolution \citep{GovWill07,GnedKravt10,AgerTey10,GovBrook10,GreifGlov10}. 

Several models were developed in the past to descibe the multi-phase ISM \citep[e.~g., ][]{McKee1977,Yepes1997,Gnedin1998,Klypin1998,Hultman1999,StinSeth06}. An often used type of model for star formation and stellar feedback in cosmological smoothed particle hydrodynamics (SPH) simulations is described in \citet{SpringHern03} (SH03), which is an adaption of the model introduced by \citet{Yepes1997}. Basically, rate equations for the densities of the cold and hot gas phases are formulated, including sources and sinks related to star formation and feedback from supernovae. Recently, a variety of phenomenological models that treat particular physical processes in the ISM have been proposed \citep[e.~g., ][]{Gnedin2009,KrumKee09,JoungLow09,MurMona10,OstKee10,PadNord09}. Some of these models are designed to account for sub-grid scale physics in numerical simulations. Others are mainly intended to obtain analytical or semi-analytical predictions that can be compared to observations. Even so, particular components of the latter class of models could be incorporated into an SGS model. In the following, we briefly review these models from the perspective of the physical processes involved.

\citet{PadNord09} [PN11] parametrize the star formation rate per free-fall time as a function of the virial parameter, i.~e., the turbulent velocity dispersion relative to the specific gravitational energy, by using data from forced isothermal MHD turbulence simulations. Following \citet{KrumKee05} [KM05], the star formation rate is calculated by integrating density fluctuations beyond a critical density that is given by the virial parameter and the Mach number of the turbulent cold neutral medium. However, as pointed out by  \citet{KrumKee09} [KMT09], new observations reveal a tight correlation between the molecular hydrogen surface density and the star formation rate. They present an analytic model that includes approximate calculations of molecular hydrogen fraction from a spherical-cloud model and the star formation efficiency per free-fall time on the basis of the numerical parametrization in KM05. This model reproduces the Kennicutt-Schmidt relation between the star formation rate and the surface density on length scales of the order of a kpc in recent surveys.

By assuming a constant star formation efficiency, the formation of molecular hydrogen in cosmological simulations is modelled by an approximate treatment of shielding and photo-dissociation in \citet{Gnedin2009} [GTK09]. As in KMT09, the star formation rate is assumed to be proportional to the molecular hydrogen density rather than the density of the cold neutral medium. The unresolved density structure of the  gas is parametrized by a clumping factor, and the efficiency of star formation per free-fall time in molecular clouds is set to $1\,\%$. Using this model, \citet{GnedKravt10} investigate the Kennicutt-Schmidt relation in galaxies at high redshifts. For simulations of isolated discs with molecular hydrogen chemistry, see \citet{DobbsGlov08,RobertKravt08}. 

The KMT09 and GTK09 models focus on molecular hydrogen to predict the star formation rate, whereas the multi-phase structure and the turbulent dynamics of the ISM are not addressed explicitly. In contrast, \citet{KoppTheis98} formulate a dynamical model for the evolution of a massive and a low-mass star component and clouds embedded in hot gas, with various interaction processes. In a similar way, the model of \citet{SpringHern03} considers interacting cold and warm phases and stars. A simple multi-phase SGS model of star formation and supernova feedback is proposed by \citet{MurMona10}. By assuming that the amount of molecular hydrogen is controlled by the pressure of the ISM, rate equations for the mass and the energy of a cold and a warm phase are solved in addition to the mass that is converted into stars. \citet{OstKee10} present a considerably more detailed analytical model that separates the ISM into a diffusive gas component and into gravitationally-bound clouds, in which stars are formed at a rate that is proportional to their mass. The basic parameters of this model are the ratio of the thermal to the effective pressure (i.~e., the sum of thermal, turbulent and magnetic pressures) of the diffusive gas, the fraction of warm diffusive gas (complementing a cold diffusive phase), and the star formation efficiency of the clouds. The main idea is that the radiation of young massive stars heats the diffusive ISM and the mass exchange between the diffusive components and the clouds regulates star formation. Turbulence and the conversion of atomic into molecular hydrogen are not decisive for the regulation process.

To include SN feedback in cosmological simulations, for example, \citet{StinSeth06} model the impact of SN blast waves on the thermal structure of the ISM. In contrast, \citet{JoungLow09} [JMB09] propose a non-thermal treatment of SN feedback. They formulate a dynamical equation to compute the numerically unresolved turbulent pressure of the ISM, with the rate of energy injection by SN blast waves as source term (\emph{internal turbulence driving}). The turbulent pressure is proportional to the energy density of numerically unresolved turbulent velocity fluctuations. The coefficients of the equation for the turbulent pressure are calibrated on galactic-scale simulations of the ISM. A similar approach is utilised in \citet{ScannBruegg10} [SB10] for turbulence in galaxy outflows.

Although feedback models using the turbulent pressure are promising, JMB09 and SB10 do not account for the increase of the turbulent pressure by the energy transfer from resolved to unresolved scales via the turbulent cascade (\emph{external turbulence driving}). We expect this production channel to be important because turbulence in the ISM is to some extent driven by gravitational instabilities on large, galactic scales \citep{RomBurk10,KlessHenne10}. For the local computation in numerical simulations, \citet{SchmFed10} [SF11] formulated and tested an SGS model for highly compressible turbulence. This model is also based on a dynamical equation for the numerically unresolved turbulent energy. However, in addition to diffusion and dissipation terms, SGS turbulent energy is produced by the shear of resolved small-scale fluctuations, i.~e., the turbulent cascade. The rate of production by the turbulent cascade is called the turbulent energy flux. Simulations of forced supersonic turbulence \citep{SchmFeder09,Federrath2010b} were used to verify a new closure for the compressible turbulent energy flux. In large eddy simulations (LES), a closure is an approximation to a quantity that depends on unresolved density and velocity fluctuations. Moreover, it is demonstrated that the SGS model fulfils several basic requirements, such as a constant mean dissipation rate, independent of the numerical resolution, and a power-law scaling of the SGS turbulent energy. For compressible turbulence driven by large-scale instabilities, this SGS model is the only model for computing the turbulent pressure consistently that has been systematically tested so far. Feedback can be included as an additional production term in the SGS turbulent energy equation. Since the unresolved turbulent velocity fluctuations in galaxy simulations are comparable to the speed of sound, we expect significant effects of the corresponding turbulent pressure, particularly with regard to the regulation of star formation.

The aim of this work is to bring together different approaches, using the SH03 model as a basic framework. Our treatment of star formation and molecular hydrogen formation is guided by KM05, KMT09, and PN11. To heat the interstellar gas, Lyman-continuum radiation of young massive stars and supernova feedback are calculated from the modelled star formation history, assuming the \citet{Chabrier2001} initial mass function. We incorporate internal turbulence driving by the thermal instability and by a non-thermal fraction of supernovae feedback, as in JMB09. By adding the turbulent pressure to the thermal pressure, turbulence influences the pressure balance between the phases and, in the highly turbulent regime, it significantly affects the gravitationally unstable mass fraction in the cold-gas phase. The key to the fluid-dynamical computation of the turbulent energy, including external driving via a turbulent cascade, is SF11. 

In this paper, we devise a semi-analytic formulation to describe the evolution of the two gas phases, turbulence, star formation, and feedback by averaged quantities in a box of given size. These one-zone calculations allow us to investigate the dependence on the control parameters (total gas density, metallicity, constant rate of turbulent energy production by external driving) and the coefficients of the models. In particular, we calculate the star formation efficiency for self-regulated equilibria. These equilibrium solutions are useful in their own right for a parametrization of the star formation efficiency in various astrophysical applications. The full implementation as a sub-grid scale model for cosmological and galaxy-scale simulations is the goal of future work.

An outline of the proposed multi-phase model will be given in Sect.~\ref{outline}, followed by detailed descriptions of the star formation model (Sect.~\ref{star_formation}) and the model equations for the mass and energy budgets of the warm and cold phases (Sect.~\ref{sec:evolution}). In Sect.~\ref{limit}, we consider limiting cases (single phase, constant star formation rate in equilibrium). To test our model, we discuss results from one-zone calculations in Sect.~\ref{sec:single_zone}, including a comparison with observations. Finally, we present our conclusions and an outlook to the application of the model in numerical simulations.

\section{Outline of the model}
\label{outline}

The base concept of this model is to split the density in a reference volume $V=l^3$ (i.e. a grid cell) into a cold and a warm phase density with separate thermal energy budgets, as used by \citet{SpringHern03}. The separation into two phases results from the cooling instability. In addition to the thermal energies of the cold and warm gas, the turbulent energy on the length scale $l$ is computed. Contrary to most star formation models that are used in contemporary numerical simulations, we determine the star formation efficiency per free-fall time scale based on local properties and processes of the turbulent multi-phase medium. To calculate the star formation efficiency, the typical length scale of cold-gas clumps embedded in the warm neutral medium and the fraction of molecular hydrogen are important parameters. The molecular hydrogen fraction, in turn, depends on the composition and the density of the gas. To close the system of equations, we assume virial equilibrium for the cold phase, which is largely dependent on the effective pressures, i.~e., the sum of thermal and turbulent pressures, of the phases. Since the turbulent pressure contribution is scale-dependent, the equilibrium also depends on the clump length scale.

\begin{table*}
\begin{minipage}{175mm}\centering
\caption{Set of model parameters and important variables.}\label{tab:varsandpars}
\begin{tabular}{ll}
\hline
Symbol & Description\\
\hline
\hline
 \multicolumn{2}{c}{main parameters}\\
\hline
$l$ & size of region \\ 
$\rho$ & total mass density \\
$u_{\rm c}$ & specific thermal energy of cold gas \\
$\Sigma$ & rate of energy injection by the turbulent cascade \\
$I_\nu$ & intensity of incident UV-radiation field \\
\hline
 \multicolumn{2}{c}{process parameters}\\
\hline
%$C_{\epsilon}$ & turbulent dissipation parameter\\
%$\gamma$ & polytropic equation of state parameter\\
$\epsilon_{\rm cc}$ & efficiency of cold phase evaporation by clump collisions\\
$\epsilon_{\rm tt}$ & efficiency of turbulence production via phase separation\\
$\epsilon_{\rm SN}$ & efficiency of turbulence production by SNe\\
$u_{\rm SN}$ & specific energy of SN-ejecta\\
$\eta$ & turbulent velocity scaling exponent\\
$b$ & compressive factor, describing the ratio of solenoidal and compressive turbulent modes\\
$f_{\rm loss}$ & fraction of mass ejected during prestellar collapse\\
$\zeta_{\rm m}$ & fraction of newly build up metals in SN-ejecta\\
$x_{\rm Lyc}$ & energy deposited in gas per absorbed Lyman continuum photon\\
\hline
\multicolumn{2}{c}{important variables}\\
\hline
$u_{\rm w}$ & specific thermal energy of warm gas \\ 
$e_{\rm t}$ & specific turbulent energy \\
$\rho_{\rm w}$ & fractional density of warm gas\\
$\rho_{\rm w,pa}$ & average density in the warm phase\\
$\rho_{\rm c}$ & fractional density of cold gas\\
$\rho_{\rm c,pa}$ & average density in the cold phase\\
$\rho_{\rm s}$ & averaged stellar mass density\\
$f_{\rm c,H_2}$ & mass fraction of shielded molecular gas in the cold phase\\
$l_{\rm c}$ & size of cold clumps \\
$\SFR$ & faction of shielded molecular gas converted into stars per respective free fall time\\
$\varepsilon_{\rm ff}$ & faction of total density converted into stars per respective free fall time\\
$Z$ & mass fraction of heavy elements \\
\hline
 \end{tabular}
 \end{minipage}
\end{table*}

In the following, quantities with subscript 'c' belong to the cold phase, those with 'w' to the warm phase, those with 's' to the star formation and those without the latter subscripts denote quantities of all the gas in the reference volume. An overview of used model specific parameters and variables is given in table \ref{tab:varsandpars}.

\subsection{Specific energy variables}

The total thermal energy density $u\rho$ can be expressed as sum of the thermal energies of the cold and warm phases:
\begin{equation}
 u\rho= u_{\rm{c}}\rho_{\rm{c}}+u_{\rm{w}}\rho_{\rm{w}},
\end{equation}

where fractional densities $\rho_{\rm{c}}$ and $\rho_{\rm{w}}$ are given by the masses
$m_{\rm{w}}$ and $m_{\rm{c}}$ in the warm and cold phases, respectively, divided by the
reference volume $V$, and $\rho=\rho_{\rm{c}}+\rho_{\rm{w}}$ is the total gas density.

The specific thermal energy of the warm phase, $u_{\rm{w}}$ is changed by radiative cooling and heating, the mixing of hot SN-ejecta and cold gas, and turbulent dissipative heating. On the other hand, we assume that $u_{\rm{c}}$, the specific thermal energy of the cold phase, has a constant value, corresponding to an average temperature $T_{\rm{c}}=50\;K$ of the cold phase. Numerical simulations suggest that the isothermal approximation is reasonable for the cold phase of the interstellar medium, because most of the gas in the cold gas is situated close to the asymptotically isothermal branch of the equilibrium curve between radiative cooling and heating \citep{SeiSchm09,AudHenne10}. To preserve energy conservation in our model, we account for any heating process that affects the cold gas by a transfer of a certain amount of cold gas to the warm phase.

Apart from the thermal energy, we assume that the gas in both phases has a certain specific turbulent energy $e_{\rm{t}}$ that corresponds to nearly isotropic random motions on length scales smaller than the size $l$ of the reference volume. An exact definition of $e_{\rm{t}}$ will be given on the basis of a decomposition of the fluid-dynamical equation in scale space.

\subsection{Density variables and effective pressure of the gas phases}
\label{sc:clump_length}

Since each phase fills only a fraction of the total volume $V$, we define the average densities
within the phases, $\rho_{\rm{c,pa}}$ and $\rho_{\rm{w,pa}}$,\footnote{Subscript 'pa' means 'phase average'} by the identities
\begin{alignat}{2}
\label{eq:cold_mass}
 m_{\rm{c}} &=\rho_{\rm{c,pa}}V_{\rm{c}} &=\rho_{\rm{c}}V,\\
\label{eq:warm_mass}
 m_{\rm{w}} &=\rho_{\rm{w,pa}}V_{\rm{w}} &=\rho_{\rm{w}}V.
\end{alignat}

where $V_{\rm{c}}$ is the volume occupied by the cold gas phase, and $V_{\rm{w}}=V-V_{\rm{c}}$.

Quantities such as the star formation rate, the molecular fraction in the cold phase and the cooling rate depend on $\rho_{\rm{w,pa}}$ and $\rho_{\rm{c,pa}}$. To determine $V_{\rm{c}}$, it would be necessary to know the structure of the two-phase medium on length scales smaller than $l$. In principle, one could parametrize the cold gas fraction $V_{\rm{c}}/V$ from small-scale simulations of thermally bistable turbulence \citep{SeiSchm09}. However, because of the high sensitivity of the thermal instability on the environment (boundary conditions, gas density, etc.), it is not obvious how to relate the parameters of such idealised simulations to the the local properties of a grid cell in large-scale simulations.

A much simpler approach is to assume that the cold gas is nearly in viral equilibrium if turbulence is accounted for and that clouds of cold gas with a characteristic scale $l_{\rm c}$ are embedded in the warm phase. The effect of turbulence can be described by an effective pressure that includes both microscopic (thermal) and macroscopic (non-thermal) motions (a precise definition will be given below). For a spherical cloud of density $\rho_{\rm{c,pa}}$ and size $l_{\rm c}$, the generalized virial theorem implies the equilibrium condition 
\begin{equation}
   \label{eq:virial}
   3P_{\rm c,eff} - {\frac{\pi }{5}G\rho_{\rm c,pa}^2l_{\rm c}^2 -
   3P_{\rm{w,eff}}}\simeq 0,
\end{equation}
where the effective pressure of the warm phase is substracted as external pressure \citep[see Sect. 14.1 in][]{Lequeux}. Since the turbulent pressure depends on the length scale, $P_{\rm c,eff}$ and $P_{\rm w,eff}$ are also functions of $l_{\rm c}$. In principle, this equation could be used to determine the length scale $l_{\rm c}$. It turns out, however, that the resulting system of equations is generally not well posed, meaning that no solutions exist for regions in the parameter space that definitely could be swept through in numerical simulations. As a consequence, either the relatively simple model with a single, characteristic length scale $l_{\rm c}$ has to be abandoned or the assumption of virial equilibrium as formulated above has to be loosened. A multi-scale model might eventually result from recent theoretical developments (P. Hennebelle, private communication). In this article, we choose the second option and investigate its consequences. Typically, structures satisfying Eq.~(\ref{eq:virial}) are not gravitationally bound. The dominant contributions come from the effective pressure, and these structures are held together by the pressure that is exerted by the surrounding warm gas. For this reason, the gravitational energy term can be neglected, and we obtain an approximate effective pressure balance: 
\begin{equation}
 \label{eq:eff_press_equilbr}
 P_{\rm{w,eff}}\stackrel{!}{=}P_{\rm c,eff},
\end{equation}

On the average, the turbulent pressure significantly contributes to the support of the cold gas against gravity. In order to connect the properties of the cold phase to the star formation rate, we assume that localized regions exist in the cold phase, where weak turbulent pressure support persists over sufficiently long periods of time so that the gas can collapse. The existence of such regions is a consequence of the intermittency of turbulence. The critical size of these regions is roughly given by the thermal Jeans length,
\begin{equation}
 \label{eq:jeans}
 \lambda_{\rm J,c}=
 c_{\rm c}\left(\frac{\pi}{\gamma G \rho_{\rm c,pa}}\right)^{1/2}=
 \left(\frac{\pi(\gamma-1)u_{\rm c}}{G\rho_{\rm c,pa}}\right)^{1/2},
\end{equation}
where $c_{\rm c}=[\gamma(\gamma-1)u_{\rm c}]^{1/2}$ is the speed of sound in the cold gas, $\gamma$ the polytropic equation of state parameter and $G$ the gravitational constant.
Thus, we define the length scale $l_{\rm c}$ by
\begin{equation}\label{eq:l_c}
 l_{\rm c}=\lambda_{\rm J,c}.
\end{equation}
The effective pressure equilibrium~(\ref{eq:eff_press_equilbr}) and the length scale~(\ref{eq:l_c}) really have a complementary meaning. While the former statistically accounts for the overall effect of turbulence, the latter specifies a typical size of locally collapsing structures in the cold phase. In a certain sense, this corresponds to the fact that molecular clouds do not collectively collapse although their mass is much greater than the thermal Jeans mass, while gravitationally unstable cores are formed locally \citep{LowKless04}.

From the effective pressure balance~(\ref{eq:eff_press_equilbr}) between the phases follows the ratio
\begin{equation}
 \label{eq:press_eq}
 	\frac{\rho_{\rm{c,pa}}}{\rho_{\rm{w,pa}}} = 
	r_{\rm w}:= \frac{\sigma_{\rm w,eff}}{\sigma_{\rm c,eff}},
\end{equation}

where $\sigma_{\rm w,eff}$ and $\sigma_{\rm c,eff}$ are functions of the internal energies $u_{\rm{w}}$ and $u_{\rm{c}}$, and the turbulent energy $e_{\rm{t}}$. Combining Eq.~(\ref{eq:cold_mass}-~\ref{eq:press_eq}), we can express the phase densities and volumes in terms of the fractional densities and the specific pressures:
\begin{alignat}{2}
	\label{eq:rho_c}
         \rho_{\rm{c,pa}}=&\,r_{\rm w}\rho_{\rm{w}}+\rho_{\rm{c}}, \quad
         V_{\rm{c}}=&\frac{\rho_{\rm{c}}V}{\rho_{\rm{c}}+r_{\rm w}\rho_{\rm{w}}},\\
	\label{eq:rho_w}
         \rho_{\rm{w,pa}}=&\,r_{\rm w}^{-1}\rho_{\rm{c}}+\rho_{\rm{w}},\quad      
         V_{\rm{w}}=&\frac{\rho_{\rm{w}}V}{\rho_{\rm{w}}+r_{\rm w}^{-1}\rho_{\rm{c}}}. 
\end{alignat}

Furthermore, we define the stellar density $\rho_{\rm{s}}$ to be the the stellar mass within the reference volume $V$ divided by that volume:
\begin{equation}
 \rho_{\rm{s}}=\frac{m_s}{V} 
\end{equation}

Numerical simulations of forced turbulence in thermally bistable gas indicate that the specific turbulent energy is nearly isotropic and uniformly distributed among the phases \cite{SeiSchm09}. Thus, the turbulent velocity dispersion within the cold gas can be related to the turbulent energy on the length scale $l$ via the power law
\begin{equation}
    \label{eq:sigma_c}
	3\sigma_{\rm c}^2 = 2e_{\rm{t}}\left(\frac{l_{\rm{c}}}{l}\right)^{2\eta}\;.
\end{equation}

The scaling exponent $\eta$ is constrained by $1/3\le\eta\le 1/2$, where the lower and upper bounds correspond to Kolmogorov and Burgers scaling, respectively. This scaling law is consistent with the observed $\sigma_{\rm c}$-scaling relation \citep[see, for example, ][]{Larson1981}. 

With the above definition, the effective pressure of the cold gas on the length scale $l_{\rm c}$ is given by (see SF11)
\begin{equation}
    \label{eq:press_eff_c}
    \begin{split}
	P_{\rm c,eff} &= \rho_{\rm{c,pa}}\sigma_{\rm c,eff} \equiv
	\rho_{\rm{c,pa}}\left(\frac{c_{\rm c}^2}{\gamma} + \sigma_{\rm c}^2\right) \\
	&= (\gamma-1)\rho_{\rm{c,pa}}u_{\rm c}\left(1+\frac{\gamma}{3}\mathcal{M}_{\rm c}^{2\eta}\right),
    \end{split}
\end{equation}
where $\mathcal{M}_{\rm c}=\sqrt{3}\sigma_{\rm c}/c_{\rm c}$ is the root mean square Mach number of turbulence in the cold phase. The turbulent pressure $P_{\rm w,eff}$ is given by an analogous expression, with cold-phase quantities replaced by the corresponding quantities in the warm phase. With these definitions, the variables $l_{\rm c}$, $\rho_{\rm{c,pa}}$, and $\rho_{\rm{w,pa}}$ can be determined solving Eqs.~(\ref{eq:rho_c}), (\ref{eq:rho_w}) and (\ref{eq:l_c}) iteratively.

\subsection{Gas composition variables}
The chemical composition of the gas is for simplicity, as we do not track individual species, identified by its mass fraction of heavy elements $Z$. Given the mass fraction of helium $Y_\odot$ at solar metallicity $Z_\odot$ and its primordial value $Y_{\rm prim}$, $Y$ at metallicity $Z$ is assumed to be $Y=Y_{\rm prim}+(Y_\odot-Y_{\rm prim})Z$. Then the total mass fraction of hydrogen $X$ is given by $X=1-Y-Z$. If the the gas is neutral, but not molecular, which is approximately true in the warm phase and in the cold phase gas outside of molecular cores, the mean molecular weight $\mu$ is given by
\begin{equation}
 \left(m_{\rm{H}}\mu\right)^{-1}= Xm_{\rm{H}}^{-1}+Ym_{\rm{He}}^{-1}+Zm_{\rm{Z}}^{-1},
\end{equation}
where $m_{\rm{H}}$ and $m_{\rm{He}}$ are the atomic masses of hydrogen and helium, respectively, and $m_{\rm{Z}}$ is the average atomic mass of the heavier elements. Within the molecular cores of the cold phase we assume the gas to be fully molecular.

\section{Star formation}
\label{star_formation}

Following KMT09, cold gas is converted into stars at a rate that depends on the mass of molecular hydrogen in the reference volume ($m_{\rm{H_2}} = f_{\rm c,H_2}\rho_{\rm c}V$, where $f_{\rm c,H_2}$ is the molecular hydrogen fraction in the cold gas phase):
\begin{equation}
  \label{eq:sf}
  \dot{\rho}_{\rm{s}}  = \frac{(1-f_{\rm loss})f_{\rm c,H_2}\rho_{\rm c}}{t_{\rm{s}}}.
\end{equation}
We define the star formation time scale $t_{\rm{s}}$ by
\begin{equation}
  \label{eq:sf_timescale}
  t_{\rm{s}}  = \frac{t_{\rm{c,ff}}}{\SFR},
\end{equation}
where the free-fall time scale in the cold gas is given by the phase-average (not the fractional) density:
\begin{equation}
  t_{\rm{c,ff}}^2 = \frac{3\pi}{32G\rho_{\rm{c,pa}}},
 \end{equation}
and $\SFR$ is the dimensionless star formation rate per free fall time $t_{\rm{c,ff}}$. Not all the mass in collapsing prestellar cores eventually ends up in a stars. A fraction $f_{\rm loss}\simeq 0.5\ldots0.7$ of mass is ejected during prestellar collapse \citep[e.g.][]{Matzner2000,HenneChab08,Chabrier2010}. We account for the mass ejection by correcting the star formation rate by the factor $(1-f_{\rm loss})$ in Eq.~(\ref{eq:sfr}).

To calculate $\SFR$, KMT09 derive a parametrization in terms of the gas column density, which reproduces important observational results from recent high-resolution surveys. These data also imply that the star formation is tightly correlated with the density of molecular hydrogen. This is the reason for including the factor $f_{\rm c,H_2}$ in Eq.~\ref{eq:sf_timescale}. On the other hand, \citet{GlovClark11} questioned a causal relationship between the star formation rate and the molecular hydrogen fraction. They argue that the observed correlation results form the necessity of effective shielding of star-forming regions from the interstellar radiation field. But this is in essence the effect that KMT09 describe with their model. For this reason, we also include the molecular hydrogen fraction as a coefficient in the expression for the star formation rate. 

KMT09 implicitly account for the turbulent energy by assuming that molecular clouds are virialized. In addition, the molecular cloud mass is determined by setting the Toomre stability parameter equal to unity. In Sect.~\ref{sc:clump_length} we determine the mean cold-gas density $\rho_{\rm{c,pa}}$ from an effective pressure balance, and we introduce a characteristic scale $l_{\rm{c}}$ that is given by the thermal Jeans mass for this density. Since turbulence in the cold phase is generally supersonic, the local density of the gas greatly fluctuates. Therefore, we consider a statistical ensemble of overdense structures on the length scale $l_{\rm{c}}$. For convenience, we call these structures \emph{clumps}. The greater the overdensity relative to $\rho_{\rm{c,pa}}$, the smaller the critical density for gravitational collapse. For a given statistical distribution of density fluctuations, which we assume to be log-normal, the dimensionless star formation rate $\SFR$ then can be calculated as proposed by PN11.
 
\subsection{Star formation efficiency}\label{sec:SFR}

By assuming a one-dimensional root mean square (rms) turbulent velocity dispersion $\sigma_{\rm c}$, the virial parameter of a clump at the mean density $\rho_{\rm{c,pa}}$ is given by \citet{BertKee92}
\begin{equation}
  \label{eq:vir_c}
  \alpha_{\rm{vir}} 
  =\frac{15}{\pi G\rho_{\rm{c,pa}}}\left(\frac{\sigma_{\rm c}}{l_{\rm{c}}}\right)^2
  \propto \left(\frac{t_{\rm{c,ff}}}{t_{\rm{c,dyn}}}\right)^2,
\end{equation}

where $t_{\rm{c,dyn}}=l_{\rm{c}}/(\sqrt{3}\,\sigma_{\rm c})$ is the dynamical time scale. Since the statistical ensemble of clumps has to encompass the whole cold phase, not only regions of weak turbulent support, the rms velocity dispersion is given by the turbulent energy scaled down to the length scale $l_{\rm{c}}$. Hence, by substituting the scaling law (Eq.~ \ref{eq:sigma_c}), the virial parameter can be expressed as
\begin{equation}
  \label{eq:vir_l}
  \alpha_{\rm{vir}} = 
  \frac{10}{\pi G\rho_{\rm{c,pa}}}\cdot
  \frac{e_{\rm{t}}}{l_{\rm{c}}^{2(1-\eta)}l^{2\eta}}.
\end{equation}

PN11 argue that the overdensity in compressed shock layers is proportional to the square of the Mach number, $\mathcal{M}_{\rm c}^2$ (see Sect.~\ref{sc:clump_length} for the definition of $\mathcal{M}_{\rm c}$). By applying the Jeans criterion for the gravitational collapse of a compressed region within the cold gas phase, it follows that the critical overdensity ratio $x_{\rm{crit}}=\rho_{\rm{c,crit}}/\rho_{\rm{c,pa}}$ is proportional to $\alpha_{\rm{vir}}\mathcal{M}^2_{\rm{c}}$:
\begin{equation}
  \label{eq:x_crit}
  \begin{split}
  x_{\rm crit} &= 0.0371\alpha_{\rm{vir}}\mathcal{M}^2_{\rm{c}}\\
  &=
  \frac{0.0742}{G\gamma(\gamma-1)\rho_{\rm{c,pa}}u_{\rm c}}\cdot
  \frac{e_{\rm{t}}^2}{l_{\rm{c}}^{2(1-2\eta)}l^{4\eta}}.
  \end{split}
\end{equation}

The constant of proportionality in the above equation is fixed by the definition of the Bonnor-Ebert radius (see PN11). For MHD turbulence, PN11 show that the $x_{\rm{crit}}$ differs by a factor $\beta$ that specifies the ratio of thermal to magnetic pressures. 

Since $u_{\rm c}=\mathrm{const.}$, the variation of the critical density
$\rho_{\rm{crit}}=x_{\rm{crit}}\rho_{\rm{c,pa}}$ is solely determined by the second factor in Eq.~(\ref{eq:x_crit}). For the two limiting cases of Kolmogorov and Burgers scaling, we obtain
\[
  \rho_{\rm crit} \propto
  \begin{cases}
    e_{\rm{t}}^{2}\,l_{\rm{c}}^{-2/3}l^{-4/3} & \mbox{if } \eta=1/3, \\
    e_{\rm{t}}^{2}\,l^{-2} & \mbox{if } \eta=1/2.
  \end{cases}
\]

If the warm phase dominates ($l_{\rm{c}}\ll l$), then $\eta\approx 1/3$ because, averaged over a region of size $l$, turbulence is mostly subsonic. From the scaling law $e_{\rm{t}}\propto l^{2/3}$, it follows that $\rho_{\rm crit} \propto l_{\rm{c}}^{-2/3}$. The scaling behaviour of the critical density follows from the steeper decrease of self-gravity with the clump size relative to the lower turbulent energy on smaller length scales. The assumption of Kolmogorov scaling is not at odds with supersonic turbulence within the clumps, because supersonic scaling applies to length scales $l\lesssim l_{\rm{c}}$ only. However, the assumption of a uniform velocity dispersion among both phases might break down for large $\mathcal{M}_{\rm c}$. On the other hand, if the cold gas phase fills most of the volume $V=l^{3}$, $\eta$ assumes a value greater than $1/3$, depending on $\mathcal{M}_{\rm c}$. In the limit of high turbulent Mach numbers, Burgers scaling ($e_{\rm{t}}\propto l$) implies that $\rho_{\rm crit}$ becomes nearly scale-invariant.\footnote{In this case, the coefficient following from the assumption of spherical clumps would not be appropriate, but the scaling remains unaffected.} In this case, however, a  potential problem is that the turbulent pressure within the overdense cores (i.~e., on the length scale of the shock-compressed layer, which is small compared to $l_{\rm{c}}$) can exceed the thermal pressure. Consequently, the model overestimates the the mass that can form star in the limit of strongly supersonic clumps of size $l_{\rm{c}}\sim l$. Applying the model in numerical simulations, it has to be ensured that this case is sufficiently rare.

As in KM05, the mass fraction per free fall time that is converted into stars is derived from the formula
\begin{equation}
  \label{eq:sfr_ff}
  \SFR=\int_{x_{\rm{crit}}}^\infty x p(x)\,{\rm d}x,
\end{equation}

where $p(x)$ is the probability density function (pdf) of the mass density, and $x=\rho_{\rm{c,loc}}/\rho_{\rm{c,pa}}$ is the ratio of the local and mean densities in the cold phase.

For isothermal gas, the probability density function (pdf) of the gas density is approximately log-normal \citep[e.g., ][]{KritNor07,FederKless08}:
\begin{equation}
 \label{eq:pdf_lognorm}
 p(x)\,{\rm d}x=\frac{x^{-1}}{\left(2\pi\sigma^2\right)^{1/2}}\,\exp\left[-\frac{\left(\ln(x)-\left\langle\ln(x)\right\rangle\right)^2}{2\sigma^2}\right]\,{\rm d}x,
\end{equation}

where $\sigma^2=-\langle\ln(x)\rangle$ is the standard deviation of logarithmic overdensity. Log-normal fits to the density pdfs from the numerical simulations suggest the following empirical relation between $\sigma$ and the sonic Mach number:
\begin{equation} \label{eq:sigmapdf}
 \sigma^2 \approx \mathrm{ln}\left(1+b^2\mathcal{M}_{\rm{c}}^2\right).
\end{equation}

As shown by \citet{Federrath2010b}, the parameter $b$ depends on the mixture of solenoidal and compressive forcing modes, which is specified by the weighing parameter $\zeta$ of the Helmholtz decomposition of the forcing modes:
\begin{equation}
 b=\frac{1}{3}+\frac{2}{3}\left(\frac{(1-\zeta)^2}{1-2\zeta+3\zeta^2}\right)^3.
\end{equation}

For solenoidal (divergence-free) forcing, $\zeta=1$. On the other hand, $\zeta=0$ for compressive (rotation-free) forcing. Substituting the log-normal pdf~(\ref{eq:pdf_lognorm}) into Eq.~(\ref{eq:sfr_ff}), the dimensionless star formation rate is given by
\begin{equation}
 \label{eq:sfr}
 \SFR=\frac{1}{2}+\frac{1}{2} \mathrm{erf} \left[\frac{\sigma^2-2\ln\left(x_{\rm{crit}}\right)}{2^{3/2}\sigma}\right].
\end{equation}

Numerical simulations of self-gravitating turbulence \citep[e.g.][]{Klessen2001,FederGlovKless2008,Cho2011,KritNor11} show changes of the high-density tail of the pdf, which affect $\SFR$.They find a power-law tail, which is associated with self-gravitating cores. Simulations by \citet{BallVaz11} suggest that in a star forming cloud the pdf only develops a powerlaw tail at high densities over periods of $\gtrsim10\mathrm{Myr}$, while the contribution of self-gravitating cores to the pdf is negligible in the earlier phase and, thus, the shape is close to log-normal. Since the model of PN11 for $\SFR$ is conceptually based on the turbulence-dominated phase, it is consistent to assume a log-normal pdf. An advanced formulation of the model might also account for the later power-law phase, but this would also require substantial modifications in the ansatz for $\SFR$. We do not consider this in the present work. 

Furthermore, we assume a distribution of clump overdensities that is determined by the global rms turbulent energy to estimate the fraction of collapsing gas in our PN11-like calculation of the star formation efficiency. This amounts to a separation of the density and velocity fluctuations. Strictly, the fraction of cold gas that can collapse should be calculated from the distributions of both the density and the turbulent velocity fluctuations. As \citet{HenneChab08} have already pointed out, however, this is far from trivial, and we do not attempt to solve this problem here. 

\subsection{Molecular hydrogen fraction}\label{sec:molfrac}

The formation of $\mathrm{H}_2$-molecules as well as their radiative destruction are mostly heating processes. Because both rates are enhanced with density, overdense regions in a clump of cool but not molecular gas may be dispersed by this heating effect, before they possibly collapse gravitationally. So knowing the fraction of molecular dominated gas in a clump, as a tracer for the fraction of gas that is not affected by effective radiation induced heating, is essential to correctly estimate the star formation rate. The fraction of molecular dominated gas in a cold clump is strongly dependent on shielding radiation, which dissociates $\mathrm{H_2}$-molecules easily, from its inner parts. Here we use a St\"omgren-like approach similar to that \citet{McKee2010} used. In low metallicity environments this approach may lead to too high molecular fraction estimates, as reaction rates are too slow to establish dissociation equilibrium on short time scales \citep{KrumGned11}. But for our purpose this is fair enough, as we do not intend to track a whole chemical network of several species. Moreover a simple chemical network model, like that of \citet{Gnedin2009}, may have weaknesses, as it particularly in the case of large $\mathrm{H}_2$-fractions, which is of particular interest when looking at star formation, overestimates further $\mathrm{H}_2$-production \citep{Milosavljevic2011}. Apart from that, this approach is not designed to compute the total fraction of molecular gas but the fraction that is molecular dominated, as we totally neglect molecular hydrogen in radiation dominated areas. Nevertheless we compare the results of this appoach to obervations in Sect.~\ref{sec:comparison}.\newline
Assuming spherical clouds with diameter $l_{\rm{c}}$, one needs to calculate the radius $l_{\rm{c,H_2}}$, at which the production rate $R_{\rm{H_2,prod}}$ of $\mathrm{H_2}$ becomes greater than its destruction rate $R_{\rm{H_2,diss}}$.

The molecular fraction of cold gas then can be expressed as the ratio of the molecular volume in a clump $\propto l_{\rm{c,H_2}}^3$ and the total volume of the clump $\propto (l_{\rm{c}}/2)^3$:
\begin{equation}
 f_{\rm c,H_2} = \left(\frac{2l_{\rm{c,H_2}}}{l_{\rm{c}}}\right)^3,
\end{equation}

where $l_{\rm{c,H_2}}$ meets the condition
 \begin{equation} \label{eq:molfrac1}
\frac{R_{\rm{H_2,prod}}}{R_{\rm{H_2,diss}}(\hat{d}_{\rm{c,H_2}})}=1,
\end{equation}

where $\hat{d}_{\rm{c,H_2}}$ is the effective shielding layer thickness $\hat{d}$, at a position inside the clump, where equation (\ref{eq:molfrac1}) is true. The $H_2$-production and -destruction rates depending on $\hat{d}$, assuming extinction of dissociating radiation of the outer regions of the cold clump is not sufficient, are given by \citep[see ][]{Gnedin2009}
 \begin{align} \label{eq:molfrac2}
  R_{\rm{H_2,prod}}=&\frac{\rho_{\rm{c,pa}}^2}{m_{\rm H}^2}C_{\rm{\rho}} X\left(\frac{Z}{Z_\odot} r_{\rm{H_2,p,s}}+Xr_{\rm{H_2,p,g}}\right)
 \end{align}

 and
 \begin{equation}
  R_{\rm{H_2,diss}}(\hat{d})=I_{\rm \nu}S_{\rm{dust}}S_{\rm{H_2}}\frac{\rho_{\rm{c,pa}}}{m_{\rm H}}r_{\rm{H_2,d}}
 \end{equation}

 respectively, where $C_{\rm{\rho}}=e^{\sigma^2}$ is the clumping factor \citep{Gnedin2009} with $\sigma$ as defined in equation (\ref{eq:sigmapdf}), $Z_\odot$ the solar metal fraction, $r_{\rm{H_2,p,s}}$ the $\mathrm{H_2}$-formation rate on dust surfaces, $r_{\rm{H_2,p,g}}$ the $H_2$-formation rate in the gaseous phase, $r_{\rm{H_2,d}}$ the radiative dissociation rate, $I_{\rm \nu}$ the intensity of the homogeneous isotropic dissociating radiation field relative to the \citet{Draine1978}-field, $S_{\rm{dust}}$ and $S_{\rm{H_2}}$ are the shielding factors due to dust and $\mathrm{H_2}$ itself (see \citet{Glover2007} or \citet{Draine1996}):
 \begin{equation}
  S_{\rm{dust}}=\exp\left(-\sigma_{\rm{dust}}\frac{Z}{Z_\odot}\rho_{\rm{c,pa}}\hat{d}\right)
 \end{equation}

 \begin{equation} 
  S_{\rm{H_2}}=\frac{1-\omega_{\rm{H_2}}}{(1+x)^2}+\frac{\omega_{\rm{H_2}}}{(1+x)^{\frac{1}{2}}}\exp\left(-\sigma_{\rm{H_2}}(1+x)^\frac{1}{2}\right)
 \end{equation}

 where
 \begin{equation}
  x = f_{\rm{c,H_2,0}}\rho_{\rm{c,pa}}\hat{d}/( m_{\rm{H}} \kappa)
 \end{equation}

 with $\kappa=5\cdot10^{14}\;cm^{-2}$ and 
 \begin{equation}\label{eq:molfrac3}
  f_{\rm{c,H_2,0}}=\max\left(f_{\rm{c,H_2,min}},R_{\rm{H_2,prod}}/R_{\rm{H_2,diss}}(\hat{d}=0)\right)
 \end{equation}

 ( $f_{\rm{c,H_2,min}}\approx10^{-5}$ is the minimum molecular fraction in radiation dominated regions of the cold phase).\newline
 In the centre of the spherical clump of diameter $l_{\rm c}$ the shielding layer has the same thickness for all directions, i.e. $\hat{d}=l_{\rm c}/2$. So if 
\begin{equation}\label{eq:molcond}
 R_{\rm{H_2,prod}}/R_{\rm{H_2,diss}}(\hat{d}=l_{\rm c}/2)\leq1
\end{equation}

 holds, there is no molecular core in the clump, and thus $f_{\rm c,H_2}=0$. Otherwise there is one, which then is assumed to effectively block all dissociating radiation, trying to pass it. For a given position of scaled distance $\lambda=2\hat{l}/l_{\rm c}$ from the centre outside the molecular core ($l_{\rm c,H_2}<\hat{l}$) the scaled effective absorption layer thickness $\delta=2\hat{d}/l_{\rm c}$ is given by the mean of the absorption layer thicknesses over all sky $O$, but the solid angle of the molecular core $S$
\begin{equation}
 \delta(\lambda)=\int_{O\setminus S(\lambda)}\!\!\delta'(\Omega)g(\Omega)\mathrm{d}\Omega\left/\int_{O\setminus S(\lambda)}\!\! g(\Omega)\mathrm{d}\Omega\right.,
\end{equation}

weighted by the fraction of transmitted radiation, which is approximated by $g=e^{-\delta'}$. The number of photons, that can possibly reach that position, is due to the cores shadow reduced by
\begin{equation}
 I_{\nu,\mathrm{\mathrm{shadow}}}(\lambda)=\frac{I_\nu}{4\pi}\int_{S(\lambda)}\!\! \mathrm{d}\Omega.
\end{equation}

At the edge of the core $\lambda_{\rm c,H_2}=2l_{\rm c,H_2}/l_{\rm c}$ half the sky is obscured (i.e. $I_{\nu,\mathrm{shadow}}(\lambda_{\rm c,H_2})=I_\nu/2$). After integrating/substituting out all angular dependencies we have
\begin{equation}
 \delta_{\rm c,H_2}(\lambda_{\rm c,H_2})=\int_{\delta_{\rm min}}^{\delta_{\rm max}}\!\!\delta'g(\delta')\mathrm{d}\delta' \left/
                                              \displaystyle\int_{\delta_{\rm min}}^{\delta_{\rm max}}\!\!g(\delta')\mathrm{d}\delta'\right.,
\end{equation}

with $\delta_{\rm min}=1-\lambda_{\rm c,H_2}$, $\delta_{\rm max}=\sqrt{1-\lambda_{\rm c,H_2}^2}$ and
\begin{equation}
 g(\delta')=4\pi\delta'^2e^{-\delta'}\left(1-\left(\frac{1+\lambda_{\rm c,H_2}^2-\delta'^2}{2\lambda_{\rm c,H_2}}\right)^2\right)^{-\frac{1}{2}}.
\end{equation}

 If the equations (\ref{eq:molfrac2}) to (\ref{eq:molfrac3}) are substituted into (\ref{eq:molfrac1}) and using $I_{\nu,\mathrm{shadow}}(\lambda_{\rm c,H_2})$ instead of $I_\nu$, one obtains a transcendent equation for $z\equiv(x+1)^\frac{1}{2}$:
 \begin{equation}\label{eq:detz}
 C=C(z)\equiv\left(\frac{1-\omega_{\rm{H_2}}}{z^4}+\frac{\omega_{\rm{H_2}}}{z}e^{-\sigma_{\rm{H_2}}z}\right)e^{-D(z^2-1)}%\left(1-E(z^2-1)\right)
 \end{equation}

 where
 \begin{equation}\label{eq:molfracpara}\begin{array}{rl}
  C \equiv& \frac{\rho_{\rm{c,pa}}C_{\rho}}{I_{\rm \nu,shadow} r_{\rm{H_2,d}} m_H}\left(\frac{Zr_{\rm{H_2,p,s}}}{Z_\odot}+Xr_{\rm{H_2,p,g}}\right)\,,\\ \\
  D \equiv& \frac{\sigma_{\rm{dust}}Z\kappa}{f_{\rm{H_2,0}} Z_\odot}\,,\\ \\
  E \equiv& \frac{2\kappa m_{\rm{H}} }{l_{\rm{c}}\rho_{\rm{c,pa}} f_{\rm{H_2,0}} }\,.
 \end{array}
 \end{equation}

 Eqn.~(\ref{eq:detz}) has a single solution for every given $C$, but only solutions in the range of $ z\in\left[1\ldots z_{\rm max}\right[$ are allowed, as $l_{\rm{c,H_2}}$ would be greater than $l_{\rm{c}}/2$ if $z<1$ and\footnote{Note, that the following case is already covered by an even more restrictive condition given in equation (\ref{eq:molcond}).} $l_{\rm{c,H_2}}\leq0$ if $z\geq z_{\rm{max}}\equiv(1+E^{-1})^\frac{1}{2}$.\newline
As $\delta_{\rm c,H_2}\!(\lambda_{\rm c,H_2})$ is bijective for $\lambda_{\rm c,H_2}\in[0\ldots 1]$, we can use its inverse $\lambda_{\rm c,H_2}\!(\delta_{\rm c,H_2})$ to compute the molecular fraction
 \begin{equation}
  f_{\rm c,H_2} = \left\{
 \begin{array}{ll}
                                                               0&\mbox{ if eq. (\ref{eq:molcond}) true,}\\
                                                               \left\langle\lambda_{\rm c,H_2}\!(1-[z^2-1]E)\right\rangle^3&\mbox{ if }1<z,\\
                                                               1&\mbox{ else.}
                                                              \end{array}
 \right. 
 \end{equation}
 
\section{Evolutionary equations}\label{sec:evolution}

\subsection{Exchange of mass between the phases}\label{sec:evolrho}

The effective growth rate of the stellar mass density is given by
\begin{equation}\label{eq:raterhoS}
  \dot{\rho}_{\rm{s,eff}} = \dot{\rho}_{\rm s}-\dot{\rho}_{\rm s,fb},
\end{equation}

where the star formation rate $\dot{\rho}_{\rm{s}}$ is defined in Sect.~\ref{star_formation}, and $\dot{\rho}_{\rm s,fb}$ is the rate at which gas is returned to the warm phase via core collapse supernovae (SNe II). 

In our model, $\dot{\rho}_{\rm s,fb}$ is determined by a convolution of the past star formation rate $\dot{\rho}_{\rm{s}}(t-t')$ and the stellar initial mass function (IMF) $\mathrm{d}N_{\rm *}/\mathrm{d}m_{\rm *}$ times the initial stellar mass $m_{\rm *}$:
\begin{equation}
 \label{eq:sn_feedback}
 \dot{\rho}_{\rm s,fb}(t) = 
 \int_{t_{\rm b}}^{t_{\rm e}}\dot{\rho}_{\rm{s}}(t-t')\,
 \frac{1}{M_{\rm *}}\frac{\mathrm{d}N_{\rm *}}{\mathrm{d}m_{\rm *}}
 \frac{\mathrm{d}m_{\rm *}}{\mathrm{d}t'}\,\mathrm{d}t',
\end{equation}

where $m_{\rm *}=m_{\rm *}(t',Z)$ is the initial mass of a star that explodes as a supernova after a lifetime $t'$. The integration boundaries $t_{\rm b}$ to $t_{\rm e}$ correspond to the lifetimes of $40M_\odot$- and $8M_\odot$-stars, respectively. The IMF is normalised by the mean initial mass per solar mass
\begin{equation}
 M_{\rm *}= \frac{1}{M_\odot}
 \int_0^\infty m_{\rm *}\frac{\mathrm{d}N_{\rm *}}{\mathrm{d}m_{\rm *}}\,\mathrm{d}m_{\rm *}.
\end{equation}

\begin{figure}
  \resizebox{\hsize}{!}{\includegraphics{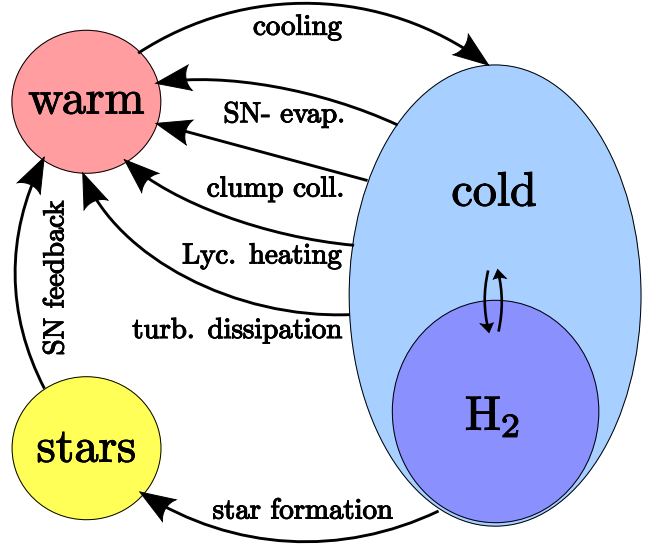}}
  \caption{Scheme of the exchange of mass. The mass budgets are depicted in yellow ($\rho_{\rm s}$), red ($\rho_{\rm w}$) and blue ($\rho_{\rm c}$), where the molecular mass ($f_{\rm c,H_2}\rho_{\rm c}$) in darker blue resides inside the cold gas mass. Arrows illustrate processes transferring mass from one to another budget.}
  \label{fig:massExchangeDiagram}
\end{figure}

For the function $m_{\rm *}(t',Z)$, we use a parametrization \citep{Raiteri1996} of the results computed by the Padova group \citep{Alongi1993,Bressan1993,Bertelli1994}. Furthermore, we assume the IMF by \citet{Chabrier2001}:
\begin{equation}
 \frac{\mathrm{d}N_{\rm *}}{\mathrm{d}m_{\rm *}}\propto\left\{\begin{array}{ll}
                                                               m_{\rm *}^{-1}e^{-\frac{\log_{10}^2\left(\frac{m_{\rm *}}{m_{\rm *,c}}\right)}{2\sigma_{\rm*,c}^2}} & \mbox{for }0.1M_\odot<m_{\rm *}<1M_\odot \\ \\
                                                               m_{\rm *}^{-2.3} & \mbox{for }1M_\odot<m_{\rm *}<125M_\odot \\ \\
                                                               0 & \mbox{ otherwise,}
                                                              \end{array}
\right.
\end{equation}

where $\sigma_{\rm*,c}=0.69$ and $m_{\rm *,c}=0.08\;M_\odot$.\\
The fraction of heavy elements in the gas increases due to SN feedback. Assuming that the metal species in the ejecta have solar relative abundances, and, that the mass fraction of newly build up metals in the ejecta of a stellar population is independent of its initial metallicity at $\zeta_{\rm m}\approx0.1$, we write
\begin{equation}\label{eq:rateZ}
\begin{split}
 \frac{\mathrm{d}(Z\rho)}{\mathrm{d}t}=&-Z\dot{\rho}_{\rm{s}}+\left(\frac{}{}\!\zeta_{\rm m}\dot{\rho}_{\rm s,fb}\right. \\
 &\left.+\int_{t_{\rm b}}^{t_{\rm e}}\!\!Z(\hat{t})\dot{\rho}_{\rm{s}}(\hat{t})\,
 \frac{1}{M_{\rm *}}\frac{\mathrm{d}N_{\rm *}}{\mathrm{d}m_{\rm *}}
 \frac{\mathrm{d}m_{\rm *}}{\mathrm{d}t'}\,\mathrm{d}t'\right),
\end{split}
\end{equation}

with $\hat{t}=t-t'$.\\
The rate of change of the fractional density of the cold phase, $\dot{\rho}_{\rm{c}}$, is determined by the processes that are described in the following. The first three processes are modelled as in \citet{SpringHern03}. For a schematic overview, see Figure~\ref{fig:massExchangeDiagram}.
\begin{enumerate}
\item Star formation reduces the mass of cold gas:
\begin{equation}
 \left.\frac{\mathrm{d}\rho_{\rm{c}}}{\mathrm{d}t}\right|_{\rm{SF}}=-\dot{\rho}_{\rm{s}}.
\end{equation}

\item Hot SN bubbles can evaporate cold clumps. Effectively, the energy that is injected by blast waves into the interstellar gas is instantaneously dissipated into heat on length scales that are much smaller than $l$. Since we cannot resolve the mixing processes, the dissipative heating and the heat conduction on these scales, we account for these processes by an evaporation rate of the cold gas,
\begin{equation}
 \left.\frac{\mathrm{d}\rho_{\rm{c}}}{\mathrm{d}t}\right|_{\rm{SN}} = A\dot{\rho}_{\rm{s,fb}},
\end{equation}

where $\dot{\rho}_{\rm{s,fb}}$ is defined by Eq.~(\ref{eq:sn_feedback}). Following the analytical model of \citet{McKee1977} for SN blast waves, the evaporation efficiency parameter $A$ is given by
\begin{equation}
 \label{eq:evap_effcn}
 A=A_0\left(\frac{\rho_{\rm{w}}}{\rho_{\rm{w,0}}}\right)^{-\frac{4}{5}}\left(\frac{l_{\rm{c}}}{l_{\rm{c,0}}}\right)^{-\frac{6}{5}}\left(\frac{V_{\rm{c}}}{V_{\rm{c,0}}}\right)^\frac{3}{5}, 
\end{equation}

where $\rho_{\rm{w}}=\rho-\rho_{\rm{c}}$, and the length scale $l_{\rm c}$ and the volume $V_{\rm c}$ of the cold clumps are defined by Eqs.~(\ref{eq:l_c}) and~(\ref{eq:rho_c}). To express variables in dimension-free form, we use the following scales:
\begin{align*}
 T_0 =\; &   T_{\rm TI} =   1.5\times 10^4\;\mathrm{K}, \\
 u_0 =\; &        k_BT_0/\mu m_H(\gamma -1), \\
 A_0 =\; &        u_{\rm{SN}}/2u_0, \\
 \rho_{\rm{0}} =\; &    \mu m_{\rm H}\times 10.0\,\mathrm{cm^{-3}},\\
 \rho_{\rm{c,0}} =\; & 10^{-3}\rho_{\rm{0}}, \\
 \rho_{\rm{w,0}} =\; & \rho_{\rm{0}}-\rho_{\rm{c,0}}, \\
 l_{\rm{c,0}} =\; & \lambda_{\rm{J,c}}(u_{\rm{c}},\rho_{\rm{c,pa,0}}).
\end{align*}

\item The cold phase gains mass from the warm phase via radiative cooling if the gas is thermally unstable: %[\textbf{WS: sign was wrong because cold gas is produced if $\Lambda< 0$}]
\begin{equation}
 \label{eq:density_cool}
 \left.\frac{\mathrm{d}\rho_{\rm{c}}}{\mathrm{d}t}\right|_{\rm{cool}}=
 \frac{(1-f_{\rm{th}})\rho_{\rm{w}}\Lambda_{\rm{eff}}}{u_{\rm{w}}-u_{\rm{c}}},
\end{equation}

where the effective cooling rate $\Lambda_{\rm eff}$ is defined below in Sect.~\ref{sec:evolenergy}, and $f_{\rm{th}}$ is the thermal stability indicator that switches on/off terms in the model equations that are related to the thermal instability:
\begin{equation}
 1-f_{\rm{th}}:=\left\{\begin{array}{ll}
                1 & \mbox{if conditions (a)-(c) are satisfied,}\\
                0 & \mbox{else.}\\
               \end{array}\right.
\end{equation}

The warm neutral gas is treated to be thermally unstable, in the meaning of currently separating into two pases due to cooling, if following conditions are met:
\begin{enumerate}
\renewcommand{\theenumi}{(\arabic{enumi})}
 \item The net effect of radiative cooling, Lyman continuum radiation, UV background radiation, and turbulent dissipative heating must decrease of the thermal energy, i.e. $\Lambda_{\rm eff}>0$.
 \item The warm gas density $\rho_{\rm w,pa}$ must exceed $0.1\,\mu m_{\rm H}\mathrm{cm}^{-3}$, since this is roughly the minimum density for thermal instability according to the equilibrium cooling curve.
 \item Furthermore the gas must be largely neutral to be thermally unstable, thus we assume an upper temperature threshold $T_{\rm TI}\equiv1.5\cdot10^4\;K$ for the cooling instability of the warm gas.

\end{enumerate}

\item Massive stars of spectral class O and B are strong emitters of radiation in the far ultraviolet. In particular the photons in range of the Lyman continuum (Lyc) deposit a significant amount of energy $x_{\rm{Lyc}}$ per photon (See Sect.~\ref{sec:SFReq}) in the gas, as they are absorbed and then reemitted as Lyman $\alpha$ (Ly$\alpha$) photons. The number of Lyc photons emitted by young, massive stars per unit volume and per unit time, $\dot{N}_{\rm Lyc,loc}$, is computed from a convolution of the past star formation rate $\dot{\rho}_{\rm s}$ and the specific emission rate of Lyc-photons $\dot{n}_{\rm Lyc}$:
\begin{equation}
 \dot{N}_{\rm Lyc,loc}(t)=
 \int_0^t \dot{\rho}_{\rm s}(t')\dot{n}_{\rm Lyc}(t-t',Z)\mathrm{d}t',
\end{equation}

For $\dot{n}_{\rm Lyc}(t-t',Z)$, we use an analytic fit to data from evolutionary synthesis models of a simple star population \citep{Kotulla2009}. Some fraction $f_{\rm leak}$ of these photons may leak into the environment of the reference volume, while $\dot{N}_{\rm Lyc,ext}$ photons from external sources may get in. The effective number of Lyc photons per unit time and unit volume that actually ionize hydrogen is then given by
\begin{equation}
 \dot{N}_{\rm Lyc}=(\dot{N}_{\rm Lyc,loc}+\dot{N}_{\rm Lyc,ext})(1-f_{\rm leak}),
\end{equation}

where $f_{\rm leak}\simeq\exp(-\sigma_{\rm H}X\rho l/m_{\rm H})$ with the ionization crossection $\sigma_{\rm H}\simeq6.3\cdot10^{-18}\;\mathrm{cm}^{-2}$. In case of average hydrogen number densities $\rho X/m_{\rm H}\geq1\;\mathrm{cm}^{-3}$ and length scales $l$ of at least a few parsec, $f_{\rm leak}$ is negligible small. It is likely that the reference volume is surrounded by an environment of comparable density and size, if sources of Lyc radiation are located there, thus $\dot{N}_{\rm Lyc,ext}$ is negligible, too. Hence, we set $\dot{N}_{\rm Lyc}=\dot{N}_{\rm Lyc,loc}$ and assume that every hydrogen atom has the same chance to absorb a photon. The radiative evaporation rate is given by:
\begin{equation}
 \left.\frac{\mathrm{d}\rho_{\rm{c}}}{\mathrm{d}t}\right|_{\rm{heat}}=
 -\frac{\dot{N}_{\rm Lyc}x_{\rm{Lyc}}\rho_{\rm{c}}}{ (u_{\rm{w}}-u_{\rm{c}})\rho}.
\end{equation}

\item Turbulent energy is dissipated into thermal energy at a rate that is given by
\begin{equation}\label{eq:turbDiss}
 \epsilon=C_\epsilon \frac{e_{\rm{t}}^{3/2}}{l},
\end{equation}

where $C_\epsilon$ is about unity (see \citealt{SchmNie06b}, SF11). In our two-phase model, turbulent dissipation heats the gas in the warm phase, but we
assume the temperature of the cold gas to be constant. To compensate turbulent dissipation in the cold phase, we simply transfer an equivalent amount of mass from the cold to the warm phase. Since turbulence is assumed to be homogeneous on length scales smaller than $l$, the amount of energy dissipated in the cold phase is $m_{\rm c}\epsilon\,\mathrm{d}t$ over an infinitesimal time interval $\mathrm{d}t$. Setting this equal to the increase of thermal energy $(u_{\rm w}-u_{\rm c})\mathrm{d}m_{\rm c}$ if the mass $\mathrm{d}m_{\rm c}$ is transferred to the warm phase and substituting $m_{\rm c}=\rho_{\rm c}V$, we obtain
\begin{equation}
  \left.\frac{\mathrm{d}\rho_{\rm{c}}}{\mathrm{d}t}\right|_{\rm diss}=
  -\frac{C_\epsilon e_{\rm t}^{3/2}\rho_{\rm{c}}}{l(u_{\rm{w}}-u_{\rm{c}})}\,.
\end{equation}

\item If the cold gas forms small compact clumps embedded in the warm phase, i.~e., $l_{\rm{c}} \ll l$, we can model interactions between the clumps as collisions. Since collisions cause a certain mass loss of the cold phase by turbulent mixing and heating, we write
\begin{equation}
 \left.\frac{\mathrm{d}\rho_{\rm{c}}}{\mathrm{d}t}\right|_{\rm{coll}}=
 -\epsilon_{\rm{cc}}\rho_{\rm{c,pa}}r_{\rm{c,coll}}l_{\rm{c}}^3.
\end{equation}

The effect of clump collisions on the cold gas density is modelled by the efficiency parameter $\epsilon_{\rm{cc}}$ and the the collision rate 
\begin{equation}
 r_{\rm{c,coll}}=n_{\rm{c}}V\frac{v_{\rm{c,coll}}}{l_{\rm{c,free}}}\,.
\end{equation}

Setting the average volume of a cold clump equal to $\pi l_{\rm{c}}^3/6$, the number density of the clumps is $n_{\rm{c}}\sim (6V_{\rm{c}}/\pi l_{\rm{c}}^3)/V$ and the mean free path $l_{\rm{c,free}}=(\pi l_{\rm{c}}^2n _{\rm{c}})^{-1}$. The rms velocity of the clump motion in the surrounding warm medium can be estimated from the square root of the turbulent energy $e_{\rm{t}}$, corrected by the internal velocity dispersion $\sigma_{\rm c}^2$ of the clumps (see Eq.~\ref{eq:sigma_c}):
\begin{equation}\label{eq:clumpvelocity}
 v_{\rm{c,coll}}=
 \left(2 e_{\rm{t}} - 3\sigma_{\rm c}^2\right)^{1/2} =
 \left[2 e_{\rm{t}}\left(1-\left(\frac{l_{\rm{c}}}{l}\right)^{2\eta}\right)\right]^{1/2},
\end{equation}

With the above definitions, it follows that
\begin{equation}
 \label{eq:coll_rate}
 r_{\rm{c,coll}}=
 \frac{36V_{c}^2}{\pi l_{\rm{c}}^4 V}
 \left[2e_{\rm{t}}\left(1-\left(\frac{l_{\rm{c}}}{l}\right)^{2\eta}\right)\right]^{1/2},
\end{equation}

where $V_{c}$ and $l_{\rm{c}}$ are given by Eqs.~(\ref{eq:rho_c}) and~(\ref{eq:l_c}), respectively.
\end{enumerate}

Collecting all six contributions, the evolutionary equation of the cold phase density can be written as
\begin{equation}\label{eq:raterhoC}
\begin{split}
  \dot{\rho}_{\rm{c}} =& 
  -\dot{\rho}_{\rm{s}}-A\dot{\rho}_{\rm{s,fb}}
  -\frac{1}{u_{\rm{w}}-u_{\rm{c}}}\\
  &\times\left[-(1-f_{\rm{th}})\rho_{\rm{w}}\Lambda_{\rm{eff}}
  +\frac{\dot{N}_{\rm{Lyc}}x_{\rm{Lyc}}\rho_{\rm{c}}}{\rho}
  +\frac{C_\epsilon e_{\rm t}^{3/2}\rho_{\rm{c}}}{l}\right]
\\
  & -\frac{36\epsilon_{\rm{cc}}\rho_{\rm{c,pa}}V_{c}^2}{\pi l_{\rm{c}}}
 \left[2e_{\rm{t}}\left(1-\left(\frac{l_{\rm{c}}}{l}\right)^{2\eta}\right)\right]^{1/2},
\end{split}
\end{equation}

and the change of the gas density in the warm phase follows from mass conservation ($\dot{\rho}_{\rm{w}}+\dot{\rho}_{\rm{c}}+\dot{\rho}_{\rm{s,eff}}=0$):
\begin{equation}\label{eq:raterhoH}
  \dot{\rho}_{\rm{w}} = -\dot{\rho}_{\rm{c}}-\dot{\rho}_{\rm{s,eff}}.
\end{equation}

\subsection{Exchange of energy between the phases}\label{sec:evolenergy}
In numerical simulations of thermally bistable turbulence \citep[e.~g., ][]{AudHenne10,SeiSchm09}, most of the gas in the cold phase is close to the isothermal equilibrium branch of the cooling curve. This is mainly caused by the higher opacity of the dense, cold gas that lowers the efficiency of radiative cooling. Other processes that affect the gas temperature in the cold phase, such as the gravitational collapse of dense regions or chemical reactions, are not explicitly treated in our model. Consequently, the specific thermal energy of the cold phase $u_{\rm{c}}$ is assumed to be constant. 
We set the temperature to a fiducial value $T_{\rm{c}}=50\mbox{ K}$, corresponding to
the lower cutoff of the cooling curve in galaxy simulations.

\begin{figure}
  \resizebox{\hsize}{!}{\includegraphics{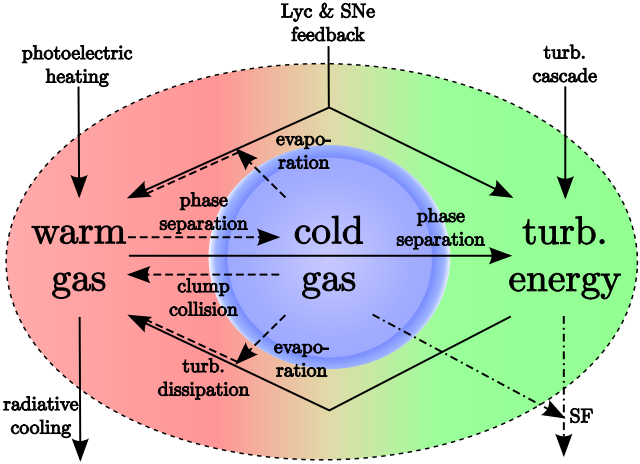}}
  \caption{Scheme of the exchange of energy. The ellipse depicts the energy content of the gas, which is separated into three budgets: the thermal energy of the warm gas $\rho_{\rm w}u_{\rm w}$ (red) and the cold gas $\rho_{\rm c}u_{\rm c}$ (blue) and the non-thermal, turbulent energy $\rho e_{\rm t}$ (green). Cold phase thermal energy can only change by loss or gain of mass, as $u_{\rm c}=const.$, this is shown by dashed arrows, while processes not intermixing cold gas with any phase have solid arrows. star formation removes mass from the cold gas, along with its energies, which is shown in dot-dashed arrows.}
  \label{fig:energyExchangeDiagram}
\end{figure}

The specific thermal energy of the warm phase, on the other hand, is changed by the processes that are discussed in Sect.~\ref{sec:evolrho}. The effects of these processes on $u_{\rm{w}}$ are as follows (see also Figure~\ref{fig:energyExchangeDiagram}).
\begin{enumerate}

 \item SNe heat the warm gas and transfer gas from the cold to the warm phase via evaporation. The energy release per unit mass is $u_{\rm SN}\approx6\cdot10^{49}\,\mathrm{erg}/M_{\rm{\odot}}$ (see Sect. ~(\ref{sec:SFReq})). Assuming that a certain fraction $\epsilon_{\rm SN}u_{\rm{SN}}$ of the feedback is non-thermal (see Sect.~\ref{sc:turb_energy}), we have
\begin{equation}
 \left.\frac{\mathrm{d}(\rho_{\rm{w}}u_{\rm{w}})}{\mathrm{d}t}\right|_{\rm{SN}}=
 [(1-\epsilon_{\rm SN})u_{\rm{SN}} + A u_{\rm{c}}]\dot{\rho}_{\rm{s,fb}}.
\end{equation}

The efficiency parameter $A$ is defined by Eq.~(\ref{eq:evap_effcn}). The rate of change of the warm gas density due to SN feedback is given by the mass ejection from SNe and the evaporation of cold gas:
\begin{equation}
 \left.\frac{\mathrm{d}\rho_{\rm{w}}}{\mathrm{d}t}\right|_{\rm{SN}} = (1+A)\dot{\rho}_{\rm{s,fb}},
\end{equation}

Combining the above equations, it follows that
\begin{equation}
 \left.\frac{\mathrm{d}u_{\rm{w}}}{\mathrm{d}t}\right|_{\rm{SN}}=
 [(1-\epsilon_{\rm SN})u_{\rm{SN}}+Au_{\rm{c}}-(1+A)u_{\rm{w}})]
 \frac{\dot{\rho}_{\rm{s,fb}}}{\rho_{\rm{w}}}.
\end{equation}

\item If the warm phase is thermally stable, the warm gas cools (or heats) at a rate given by its effective cooling function $\Lambda_{\rm{eff}}$. Once the thermal instability sets in, gas in the warm phase is converted into cold gas without changing the temperature of the remaining warm gas. We also assume that the cooling instability produces turbulent energy with an efficiency $\epsilon_{\rm{tt}}$ relative to the cooling function. Consequently, the total change of the internal energy density of the warm phase can be written as
\begin{equation}
 \begin{split}
 \left.\frac{\mathrm{d}(\rho_{\rm{w}}u_{\rm{w}})}{\mathrm{d}t}\right|_{\rm{cool}}= &\;
 -\rho_{\rm{w}}\Lambda_{\rm{eff}} 
 +u_{\rm{c}}\left.\frac{\mathrm{d}\rho_{\rm{w}}}{\mathrm{d}t}\right|_{\rm{cool}} \\
 &-(1-f_{\rm{th}})\epsilon_{\rm{tt}}\rho_{\rm{w}}\Lambda_{\rm{eff}}.
 \end{split}
\end{equation}

Since the rate of change of the warm gas density due to cooling is given by Eq.~(\ref{eq:density_cool}) multiplied by minus one, we obtain
\begin{equation}
\left.\frac{\mathrm{d}u_{\rm{w}}}{\mathrm{d}t}\right|_{\rm{cool}}=
-f_{\rm{th}}\Lambda_{\rm{eff}}-(1-f_{\rm{th}})\epsilon_{\rm{tt}}\Lambda_{\rm{eff}},
\end{equation}

The effective cooling rate $\Lambda_{\rm{eff}}$ is defined by
\begin{equation}
\label{eq:lambda_eff}
 \Lambda_{\rm{eff}}=\Lambda_{\rm{rad}}-\Gamma_{\rm{PAH}}-\Gamma_{\rm{Lyc}}-\epsilon,
\end{equation}

where $\Lambda_{\rm{rad}}$ is the specific radiative cooling rate. In this model, we use a tabled atomic cooling function, computed using the photo-ionisation package Cloudy (version 08.00), last described by \citet{Ferland1998}, without considering molecules or dust. $\Gamma_{\rm{PAH}}$ is the photo-electric heating rate \citep{Wolfire1995} due to the external radiation field $I_\nu$ modified by a factor of $Z/Z_\odot$, and $\epsilon$ is the turbulent dissipation rate per unit mass~(\ref{eq:turbDiss}). The volume rate of heating by Lyc photons is given by
\begin{equation}
 \left.\frac{\mathrm{d}(\rho_{\rm{w}}u_{\rm{w}})}{\mathrm{d}t}\right|_{\rm{heat}}=
 \dot{N}_{\rm Lyc}x_{\rm{Lyc}}+u_{\rm{c}}\left.\frac{\mathrm{d}\rho_{\rm{w}}}{\mathrm{d}t}\right|_{\rm{heat}}.
\end{equation}

Hence, the specific heating rate is
\begin{equation}
  \Gamma_{\rm{Lyc}}=\left.\frac{\mathrm{d}u_{\rm{w}}}{\mathrm{d}t}\right|_{\rm{heat}}=
 \frac{\dot{N}_{\rm{Lyc}}x_{\rm{Lyc}}}{\rho}\,.
\end{equation}

\item Since cold gas is transferred to the warm phase by clump collisions, we have
\begin{equation}
\left.\frac{\mathrm{d}(\rho_{\rm{w}}u_{\rm{w}})}{\mathrm{d}t}\right|_{\rm{coll}}=u_{\rm{c}}\left.\frac{\mathrm{d}\rho_{\rm{c}}}{\mathrm{d}t}\right|_{\rm{coll}}.
\end{equation}

The corresponding rate of change of the specific energy is given by (see Eq.~
\begin{equation}
\left.\frac{\mathrm{d}u_{\rm{w}}}{\mathrm{d}t}\right|_{\rm{coll}} =
-\epsilon_{\rm{cc}}(u_{\rm{w}}-u_{\rm{c}}) \frac{\rho_{\rm{c,pa}}r_{\rm{c,coll}}l_{\rm{c}}^3}{\rho_{\rm{w}}},
\end{equation}

where $\epsilon_{\rm{cc}}$ is the efficiency parameter of the collisions, and the collision rate $r_{\rm{c,coll}}$ is defined by Eq.~(\ref{eq:coll_rate}).
\end{enumerate}

Adding up the contributions (i) to (v), the dynamical equation for the thermal energy of the warm phase becomes
\begin{equation}\label{eq:rateuH}
 \begin{split}
 \dot{u}_{\rm w} =\; & 
 [(1-\epsilon_{\rm SN})u_{\rm{SN}}+Au_{\rm{c}}-(1+A)u_{\rm{w}})]
 \frac{\dot{\rho}_{\rm{s,fb}}}{\rho_{\rm{w}}}\\
 &-[f_{\rm{th}}+(1-f_{\rm{th}})\epsilon_{\rm{tt}}]\Lambda_{\rm{eff}}\\
 &-\epsilon_{\rm{cc}}(u_{\rm{w}}-u_{\rm{c}}) \frac{\rho_{\rm{c,pa}}r_{\rm{c,coll}}l_{\rm{c}}^3}{\rho_{\rm{w}}}.            
 \end{split}
\end{equation}

\subsection{Turbulent energy production and dissipation}
\label{sc:turb_energy}
To formulate an equation for the turbulent energy, we assume that energy is injected at constant rate $\Sigma$ by large-scale forcing. The rate of energy injection determines
the flux of kinetic energy that is transported through the turbulent cascade from larger to smaller scales. For purely hydrodynamic isotropic turbulence, the energy flux is independent of the length scale and equal to the dissipation rate $\epsilon$ in statistical equilibrium. Applying the method of (adaptively refined) large eddy simulations, $\Sigma$ can be computed from the SF11 closure for the compressible turbulent energy cascade. For the one-zone formulation of our model, we simply express $\Sigma$ in terms of the typical magnitude of the turbulent velocity fluctuations $\mathcal{V}$ induced by the turbulent cascade on the length scale $l$:
\begin{equation}
    \label{eq:sigma}
	\Sigma = C_{\epsilon}\rho\frac{\mathcal{V}^{3/2}}{l}.
\end{equation}

For pure hydrodynamical turbulence, $\Sigma=\epsilon$ and $e_{\rm t}=0.5\mathcal{V}^2$ in equilibrium.

Neglecting turbulent diffusion and collecting the terms that exchange energy between the gas phases and turbulence, the following rate equation for the turbulent energy results:
\begin{equation}\label{eq:rateeSGS}
 \begin{split}
 \dot{e}_{\rm t} =\;& 
 (\epsilon_{\rm SN}u_{\rm{SN}}-e_{\rm t})\frac{\dot{\rho}_{\rm{s,fb}}}{\rho}
 +(1-f_{\rm{th}})\epsilon_{\rm{tt}}\Lambda_{\rm{eff}}\frac{\rho_{\rm w}}{\rho}\\
 &+\frac{\Sigma}{\rho}-C_\epsilon\frac{e_{\rm{t}}^{3/2}}{l}
 \end{split}
\end{equation}

The three sources of turbulent energy production are SN feedback on length scales comparable to $l$, the cooling instability and the turbulent energy cascade. The two efficiency parameters $\epsilon_{\rm SN}$ and $\epsilon_{\rm{tt}}$ determine the coupling of the unresolved processes to the turbulent energy. In numerical simulations, in which $l$ corresponds to the grid scale, these parameters have to be chosen appropriately. The crucial problem is that, in contrast to the cascade of turbulent eddies in the inertial sub-range of isotropic turbulence, SN feedback and the cooling instability are not self-similar. We can only assume that particular efficiency parameters apply to certain ranges of scales. One option is to use small-scale simulations of the interaction of SNe blast waves with the interstellar medium and thermally bistable flows in periodic boxes to estimate these parameters. On the other hand, the model can be calibrated \emph{a posteriori} in large-scale simulations such that observational constraints are met. We also show in Sect.~\ref{sec:single_zone}, that one-zone calculations can be utilised to find reasonable choices of the efficiency parameters. 

Combining Eqs.~(\ref{eq:raterhoC},\ref{eq:raterhoH},\ref{eq:rateuH},\ref{eq:rateeSGS}) and $u_{\rm c}=\mathrm{const.}$, the following law of energy conservation equation is obtained:  
\begin{equation}
\label{eq:rateetot}
 \begin{split}
 \frac{\mathrm{d}(\rho (u+e_{\rm t}))}{\mathrm{d}t}  =\;& 
 -\dot{\rho}_{\rm s}(u_{\rm{c}}+e_{\rm t}) + \dot{\rho}_{\rm{s,fb}}u_{\rm{SN}} \\
 &- \rho_{\rm{w}}\Lambda_{\rm{rad}}+\rho_{\rm{w}}\Gamma_{\rm PAH}\\
 & +\dot{N}_{\rm{Lyc}}x_{\rm{Lyc}}  
 + \Sigma.
 \end{split}
\end{equation}

\section{Limiting cases}
\label{limit}

\subsection{One-phase medium}
\label{sec:onephase}

If most of the gas cools down to temperatures close to $T_{\mathrm{c}}$, the separation between a dynamic warm phase and a star-forming cold phase breaks down. Assuming that the warm gas can also form stars if $u_{\rm w}\le u_{\rm w, min}$, where $u_{\rm w, min}\ge u_{\rm c}$, the star formation law~(\ref{eq:sf}) becomes
\begin{equation}
  \label{eq:sf_single}
  \dot{\rho}_{\rm{s}} = \frac{\rho}{t_{\rm{s}}},
\end{equation}

where $\rho_{\rm c,pa}=\rho$ and $\dot{\rho} = -\dot{\rho}_{\rm{s,eff}}$. In this article, our focus is on a statistical description. By choosing a length scale $l\gg l_{\rm c}=\lambda_{\rm J}(\rho)$, the star formation time scale $t_{\rm{s}}$ can be calculated as in Sect.~\ref{star_formation}. When applying this model as an SGS model in numerical simulations, however, the Truelove criterion requires $l\le l_{\rm J}(\rho)/4$. Consequently, an alternative parametrization of the star formation time scale has to be applied in the one-phase limit. This is left for future work. Practically, this case will occur in simulations of individual galaxies with high resolution, but usually not in cosmological simulations, where cold-gas clumps are sufficiently below the resolution limit.

Since the cooling instability vanishes, and the exchange of energy between the phases as well as the collisions of cold clumps become meaningless, the equation for the specific thermal energy becomes
\begin{equation}
 \dot{u}_{\rm w} \simeq\;
 [(1-\epsilon_{\rm SN})u_{\rm{SN}}-u_{\rm c})]
 \frac{\dot{\rho}_{\rm{s,fb}}}{\rho}-\Lambda_{\rm{eff}}.
\end{equation}

The subtraction of $u_{\rm c}$ in the factor that is multiplied with the feedback rate results from the removal of the energy in the gas that forms stars from the gas energy budget (see Eq.~\ref{eq:rateetot}). Practically, we can neglect the difference because $u_{\rm{SN}}\gg u_{\rm c}$. The thermal energy equation is complemented by the simplified turbulent
energy equation:
\begin{equation}
\label{eq:rateeSGS_single}
 \dot{e}_{\rm t} =
 \frac{1}{\rho}\left(\epsilon_{\rm SN}u_{\rm{SN}}\dot{\rho}_{\rm{s,fb}}+\Sigma\right)
 -C_\epsilon\frac{e_{\rm{t}}^{3/2}}{l}
\end{equation}

It is instructive to consider the asymptotic limit $u_{\rm w}\simeq u_{\rm c}=\mathrm{const.}$ Then we can set  $\dot{u}_{\rm w}\simeq 0$. For net heating ($\Lambda_{\rm{eff}}<0$), this equation cannot be fulfilled because both terms are positive. If $\Lambda_{\rm{eff}}>0$, on the other hand, the feedback rate is approximately given by a balance between thermal heating by SNe and turbulence production by cooling:
\begin{equation}
  \dot{\rho}_{\rm{s,fb}} \simeq\,\frac{\rho\Lambda_{\rm{eff}}}{(1-\epsilon_{\rm SN})u_{\rm{SN}}}.
\end{equation}

Since the $\dot{\rho}_{\rm{s,fb}}$ is related to the star formation rate~(\ref{eq:sf_single}) via Eq.~(\ref{eq:sn_feedback}), the above equation imposes a condition
on the effective cooling rate so that $u_{\rm w}\simeq u_{\rm c}=\mathrm{const.}$
 
\subsection{Equilibrium solutions}
\label{sc:equilbr}

Of particular interest is the case of self-regulation, for which the star formation rate is low and nearly constant: $\dot{\rho}_{\rm{s}}\simeq \dot{\rho}_{\rm{s,eq}}=\mathrm{const.}$ A low star formation rate means that changes in the gas density are negligible in first-order approximation. In addition, we assume that the temperature of the warm phase and the specific turbulent energy are approximately constant in the self-regulated regime and that the cooling instability is active ($f=0$). For simplicity, we neglect clump collisions. Setting $\dot{u}_{\rm w}\simeq 0$ and $\dot{e}_{\rm t}\simeq 0$ in Eqs.~(\ref{eq:rateuH}) and~(\ref{eq:rateeSGS}), respectively, it follows that
\begin{align}
 \label{eq:rateuHeq}
 [(1-\epsilon_{\rm SN})u_{\rm{SN}}+Au_{\rm{c}}-(1+A)u_{\rm{w,eq}})]
 \frac{\dot{\rho}_{\rm{s,fb}}}{\rho_{\rm{w}}}
 -\epsilon_{\rm{tt}}\Lambda_{\rm{eff}} & \simeq 0 ,\\        
 \label{eq:rateeSGSeq}
 (\epsilon_{\rm SN}u_{\rm{SN}}-e_{\rm t,eq})\frac{\dot{\rho}_{\rm{s,fb}}}{\rho}
 +\epsilon_{\rm{tt}}\Lambda_{\rm{eff}}\frac{\rho_{\rm w}}{\rho}
 +\frac{\Sigma}{\rho}-C_\epsilon\frac{e_{\rm{t,eq}}^{3/2}}{l} & \simeq 0,
\end{align}

where $u_{\rm{w,eq}}$ and ${e}_{\rm t,eq}$ are the equilibrium values.

Equation~(\ref{eq:rateeSGSeq}) imposes a condition on the feedback rate. By substituting the effective cooling rate~(\ref{eq:lambda_eff}) for $\Lambda_{\rm{eff}}$, we obtain
\begin{equation}
\begin{split}
 (\epsilon_{\rm SN}u_{\rm{SN}}-e_{\rm t,eq})\frac{\dot{\rho}_{\rm{s,fb}}}{\rho} \simeq\; &
 C_\epsilon(\rho + \epsilon_{\rm{tt}}\rho_{\rm w})\frac{e_{\rm{t,eq}}^{3/2}}{l} -\Sigma\\
 &+\epsilon_{\rm{tt}}\rho_{\rm w}(\Gamma_{\rm{PAH}}+\Gamma_{\rm{Lyc}}-\Lambda_{\rm{rad}}).
\end{split}
\end{equation}

For any reasonable choice of parameters, $e_{\rm t,eq}\ll \epsilon_{\rm SN}u_{\rm{SN}}$. Consequently, a solution exits only if
\[
  C_\epsilon(\rho + \epsilon_{\rm{tt}}\rho_{\rm w})\frac{e_{\rm{t,eq}}^{3/2}}{l} \ge
 \Sigma +\epsilon_{\rm{tt}}\rho_{\rm w}(\Lambda_{\rm{rad}}-\Gamma_{\rm{PAH}}-\Gamma_{\rm{Lyc}}).
\]
The right-hand side is always positive if the cooling instability is active, because $\Lambda_{\rm{rad}} - \Gamma_{\rm{PAH}} - \Gamma_{\rm{Lyc}} \ge \Lambda_{\rm{eff}} > 0$. For this reason, the above constraint implies that there must be a minimal turbulent energy for self-regulation:
\[
  \min e_{\rm{t,eq}} =
 \left(\frac{[\Sigma +\epsilon_{\rm{tt}}\rho_{\rm w}(\Lambda_{\rm{rad}} - \Gamma_{\rm{PAH}} - \Gamma_{\rm{Lyc}})]l}{C_\epsilon(\rho + \epsilon_{\rm{tt}}\rho_{\rm w})}\right)^{2/3}.
\]
This an important implication of the multi-phase model.

The stellar feedback rate is also a constant in the self-regulated regime. This follows immediately from Eq.~(\ref{eq:sn_feedback}) for a constant star formation rate. Then
\begin{equation}
  \label{eq:rate_fb}
  \dot{\rho}_{\rm{s,fb}}\simeq \beta\dot{\rho}_{\rm{s,eq}},
\end{equation}

where
\begin{equation}\label{eq:beta}
 \beta =
 \int_{t_{\rm b}}^{t_{\rm e}}
 \frac{1}{M_{\rm *}}\frac{\mathrm{d}N_{\rm *}}{\mathrm{d}m_{\rm *}}
 \frac{\mathrm{d}m_{\rm *}}{\mathrm{d}t'}\,\mathrm{d}t'.
\end{equation}

In principle, one could invert the equations for a given star formation rate, $\dot{\rho}_{\rm{s,eq}}$, and the effective pressure equilibrium between the phases, to obtain the cold and warm-gas densities. Due to the high non-linearity of these equations, particularly the molecular hydrogen fraction, this is very difficult in practice. It is easier to search for equilibrium solutions by computing the full set of rate equations and identifying solutions that are close to equilibrium values satisfying Eqs.~(\ref{eq:rateuHeq}), (\ref{eq:rateeSGSeq}), and~(\ref{eq:rate_fb}) in certain time intervals.

By substituting Eq.~(\ref{eq:rateeSGSeq}) into Eq.~(\ref{eq:rateuHeq}), the following expression for the equilibrium energy of the warm phase can be obtained:
\begin{equation}
\label{eq:uw_eq}
\begin{split}
  u_{\rm{w,eq}} \simeq & 
  \frac{u_{\rm{SN}}}{1+A} + \frac{A}{1+A}u_{\rm{c}} - \frac{1}{1+A}e_{\rm{t,eq}}\\
  &- \frac{1}{(1+A)\beta\dot{\rho}_{\rm{s,eq}}}
  \left(C_\epsilon\rho\frac{e_{\rm{t,eq}}^{3/2}}{l}-\Sigma\right).
\end{split}
\end{equation}

For $\Sigma=0$ (no turbulence feeding by instabilities on length scales greater than $l$) and $e_{\rm{t,eq}}=0$ (turbulent energy is neglected), the SH03 equilibrium solution for $u_{\rm{w,eq}}$ results, with the exception of the factor $A/(1+A)$. Since $A\gg 1$, however, this factor is very close to unity and the result is practically the same. In our model, the cooling instability produces turbulent energy on top of the turbulent cascade. Thus, $e_{\rm{t,eq}}>(\Sigma l/C_\epsilon\rho)^{2/3}$ (see Eq.~\ref{eq:rateuHeq}) and
\begin{equation}
\label{eq:u_SH}
  u_{\rm{w,eq}} < u_{\rm SH}:= \frac{u_{\rm{SN}}}{1+A} + u_{\rm{c}}.
\end{equation}

As a consequence, we expect that the temperature of warm gas close to equilibrium decreases in numerical simulations with a turbulence SGS model, because a fraction of the energy is in non-thermal form.

\section{Modelling the evolution of a single zone}\label{sec:single_zone}

The set of six coupled nonlinear differential equations (\ref{eq:raterhoS}, \ref{eq:raterhoC}, \ref{eq:raterhoH}, \ref{eq:rateuH}, \ref{eq:rateeSGS}, \ref{eq:rateZ}), as defined in Sect.~\ref{sec:evolution}, describe how the gas in the reference volume evolves with time. By numerically integrating the model equations over closed boxes (i.~e., single zones), we obtain statistical models for a wide range of initial conditions and parameters. These models also allow us to find equilibrium states, for which the star formation rate is small and nearly const. (see Sect.~\ref{sc:equilbr}). To characterise the star formation rate, we define a dimensionless star formation efficiency by
\begin{equation}
	\label{eq:sf_efficiency}
	\varepsilon_{\rm ff}=\frac{\dot{\rho}_{\rm s}t_{\rm ff}}{\rho}.
\end{equation}

This is the fraction of the total gas mass in the reference volume that is turned into stars over a free fall timescale $t_{\rm{ff}} = \sqrt{3\pi/32G\rho}$. It is to be distinguished from $\SFR$, which is sometimes also called star formation efficiency. However, $\SFR$ specifies the fraction of cold molecular gas converted per free fall time (see Sect.~\ref{star_formation}).

If not stated otherwise, the following standard model parameters are used: 
\begin{alignat*}{2}
\eta &= 1/3, & \qquad b &= 2/3, \\
\epsilon_{\rm tt}&=0.025, & \qquad \epsilon_{\rm SN}&=0.085,\\
\epsilon_{\rm cc}&=0.0, & x_{\rm Lyc}&= 0.1 \mbox{ eV}, \\
\qquad l&=15\mbox{ pc}, & f_{\rm loss}&= 0.6.
\end{alignat*}

We comment on the choice of these units in Sect.~\ref{sec:SFReq}. Furthermore, metal enrichment is turned off by setting $\zeta_{\rm m}=0$. We specify the strength of the turbulent energy injection $\Sigma$ by relating the velocity scale $V$ (see Eq.~\ref{eq:sigma}) to the speed of sound at the temperature $T_{\rm TI}=1.5\cdot 10^4\;\mathrm{K}$, corresponding to the maximum thermal energy $u_{\rm TI}$ of thermally unstable gas. This results in the Mach number
\begin{equation}
 \label{eq:mach_sigma}
 \mathcal{M}_{\rm \Sigma}=\frac{V}{c_{\rm TI}}=
 \left(\frac{2}{\gamma(\gamma-1)u_{\rm TI}}\right)^{1/2}\left(\frac{\Sigma l}{C_\epsilon\rho}\right)^{1/3}
\end{equation}
as a basic parameter for the external (large-scale) energy injection relative to the maximal turbulence production by the thermal instability. We use the standard value $\mathcal{M}_{\rm \Sigma}=0.2$.

In the following sections \ref{sec:fb_sequence} and \ref{eq:dyn_evol} sample evolutions of gas are discussed. The gas, as well as the formed stars, in these evolutions are confined to the reference volume, i.e. nothing but energy enters or leaves the volume. If the reference volume was embedded in a more realistic, inhomogenous, dynamic environment, like in an isolated disk galaxy simulation, it would exchange mass with its environment, as gas pressure is subject to strong variation, depending on the enviroment this may lead to substantial in- or outflows, and stars may leave the area due to their drift. This is beyond the scope of a single zone model, but to demonstrate the effects of basic model parameters it is still useful. For the equilibrium solutions discussed in Sect.~\ref{sec:SFReq} however the latter is no objection. 
\subsection{Feedback sequence}\label{sec:fb_sequence}

\begin{figure}
\centering
  \resizebox{1.0\hsize}{!}{\includegraphics{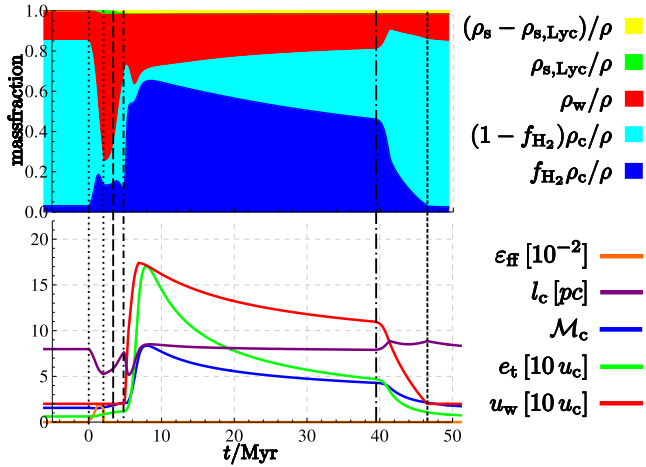}}
\caption{Demonstration of the feedback sequence: Star formation is only active for $t\in[0\ldots1]\mbox{ Myr}$ (period between the dotted vertical lines) and set to zero at later time. The upper panel shows the mass fractions of the different phases as area fillings (stellar fraction  $\rho_{\rm s}/\rho$, contribution of the young Lyc-emitting population
 $\rho_{\rm s,Lyc}/\rho$, warm fraction $\rho_{\rm w}/\rho$, neutral cold fraction $\rho_{\rm c}(1-f_{\rm c,H_2})/\rho$, and molecular fraction $f_{\rm c,H_2}\rho_{\rm c}/\rho$). In the lower panel, the evolution of the star formation efficiency $\varepsilon_{\rm ff}$, the rms Mach number of motions in the cold phase $\mathcal{M}_{\rm c}$, the size of cold clumps $l_{\rm c}$, the thermal energy of the warm phase $u_{\rm w}$, and the turbulent energy $e_{\rm t}$ are plotted.}
\label{fig:SFtrunc_evol}
\end{figure}

As described in Sects.~\ref{sec:evolrho} and~\ref{sec:evolenergy}, stellar feedback is determined by the evolution of the stellar populations within the reference volume (e.g., SN feedback starts when the first SNe II light up). As an example, we consider the evolution of gas with number density $n=\rho/(\mu m_{\rm H})=75\;\mathrm{cm}^{-3}$, solar metallicity, and no turbulent energy injection (i.e., $\Sigma=0$). To single out the effects of the stellar feedback, we artificially suppress star formation after one Myr has passed. The results are shown in Fig.~\ref{fig:SFtrunc_evol}. The evolution starts in an equilibrium between heating and cooling without star formation. Then star formation is activated within the interval indicated by the vertical dotted lines. Lyc-heating by the newly formed stars quickly lowers the star formation efficiency $\varepsilon_{\rm ff}$ defined by Eq.~(\ref{eq:sf_efficiency}), as the the cold phase gas fraction decreases, $l_{\rm c}$ drops while $f_{\rm c,H_2}$ grows, and the warm-gas pressure rises. But the specific energy $u_{\rm w}$ remains low because radiative cooling dominates in the warm phase. When Lyc-heating begins to decrease (vertical long-dashed line), more warm gas can cool down and $f_{\rm c,H_2}$ drops due to the shift in the pressure equilibrium between the phases. After the start of SN feedback (short dot-dashed vertical line) $u_{\rm w}$ and $\rho_{\rm w}$ rise quickly, and the ensuing compression of the cold phase results in an increase of $f_{\rm c,H_2}$. The kink in the evolution of $l_{\rm c}$ and $\mathcal{M}_{\rm c}$ that can be seen shortly after the onset of feedback is caused by the production of turbulent energy, which is slightly lagging behind. As SN feedback becomes weaker, $u_{\rm w}$, $e_{\rm t}$ and $f_{\rm c,H_2}$ are declining. Once all SN-progenitors are gone, SN feedback ceases (long dot dashed vertical line). Then the warm gas cools down to the initial temperature (reaching it at the short dashed line), and all quantities and gas fractions evolve back to the values for the equilibrium of heating and cooling without star formation.

\begin{figure*}\begin{minipage}{177mm}
  \resizebox{\hsize}{!}{\includegraphics{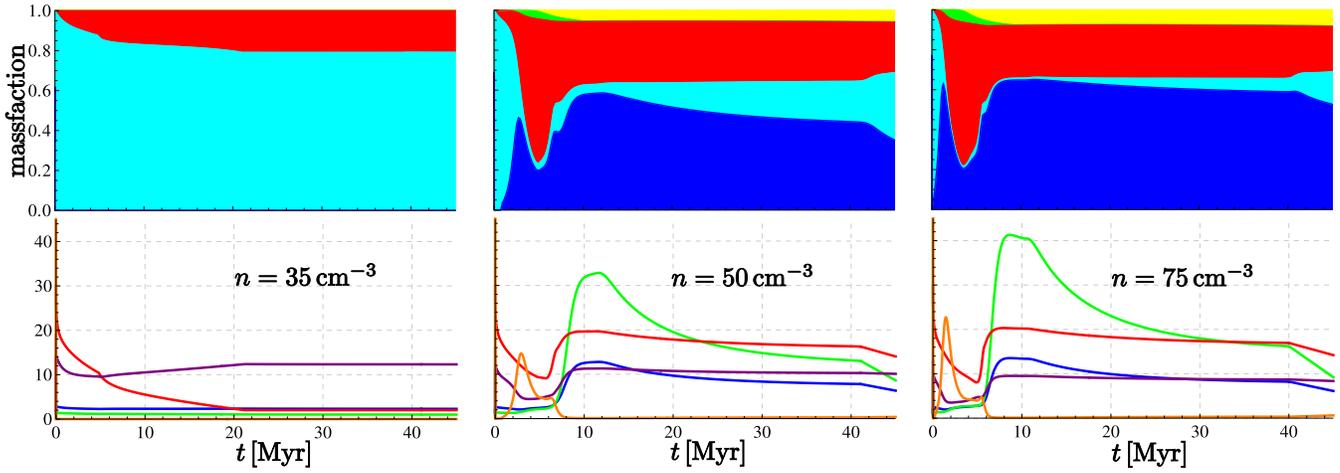}}
\caption{One-zone models for different initial gas densities at solar metallicity an moderate external driving $\mathcal{M}_\Sigma=0.2$. See the legend in Fig.~\ref{fig:SFtrunc_evol} for a definition of the plotted quantities.}
\label{fig:ND_evolution}
 \end{minipage}\end{figure*}

\subsection{Dynamical evolution}
\label{eq:dyn_evol}

The main parameters of our model are the initial total gas density $n$, the metallicity $Z$ and the forcing Mach number $\mathcal{M}_{\rm \Sigma}$, corresponding to the turbulent energy injection rate $\Sigma$. In this section, we describe the dependence of the phase evolution and star formation history on these parameters. In addition to $\rho$, $Z$, and $\mathcal{M}_{\rm \Sigma}$, several coefficients determine the relative contributions from unresolved processes. The influence of these coefficients is discussed in Sect.~\ref{sec:SFReq}.

Three sample evolutions are plotted in Fig.~\ref{fig:ND_evolution} for the initial number densities $n=35,\;50$ and $75\;\mathrm{cm}^{-3}$, $Z=Z_{\odot}$, and $\mathcal{M}_{\rm \Sigma}=0.2$. For the lowest density, no star formation occurs at all, except for a negligible fraction in the very beginning (the spurious SN feedback produced by these stars causes the kink in the cooling curve of $u_{\rm w}$ at $t\simeq4.2\mbox{ Myr}$). The final $\rho_{\rm c}/\rho_{\rm w}$ is determined by the equilibrium between turbulent dissipation, photoelectric heating and radiative cooling. The final $e_{\rm t}$ is fixed by the equilibrium between the production of turbulent energy by the thermal instability and large-scale injection and turbulent dissipation. For higher values of $n$ (middle and right panel of Fig.~\ref{fig:ND_evolution}), on the other hand, a markedly different evolution can be seen. Following an initial transient phase that ends after about 10 Myr, a stationary mode of star formation is entered, in which the star formation efficiency $\varepsilon_{\rm ff}$ is a few per mil. During the transient phase, there are three more or less distinct maxima of $\varepsilon_{\rm ff}$. This behaviour can be understood as follows. The initial rise of $\rho_{\rm w}/\rho$ due to Lyc-heating by the first stars causes a compression of the cold gas. This results in an increasing molecular hydrogen fraction and, thus, an enhancement of $\varepsilon_{\rm ff}$. Depletion of the cold phase reverses this trend. As Lyc-heating fades out, the star formation efficiency increases again, resulting in a second, but weaker peak. Due to the cooling of the warm phase, which exerts pressure on the cold phase, $\rho_{c,pa}$ is declining and $\varepsilon_{\rm ff}$ is lowered. The first SNe raise $u_{\rm w}$ and the subsequent growth of turbulent energy causes the third maximum.

\begin{figure*}\begin{minipage}{177mm}
  \resizebox{\hsize}{!}{\includegraphics{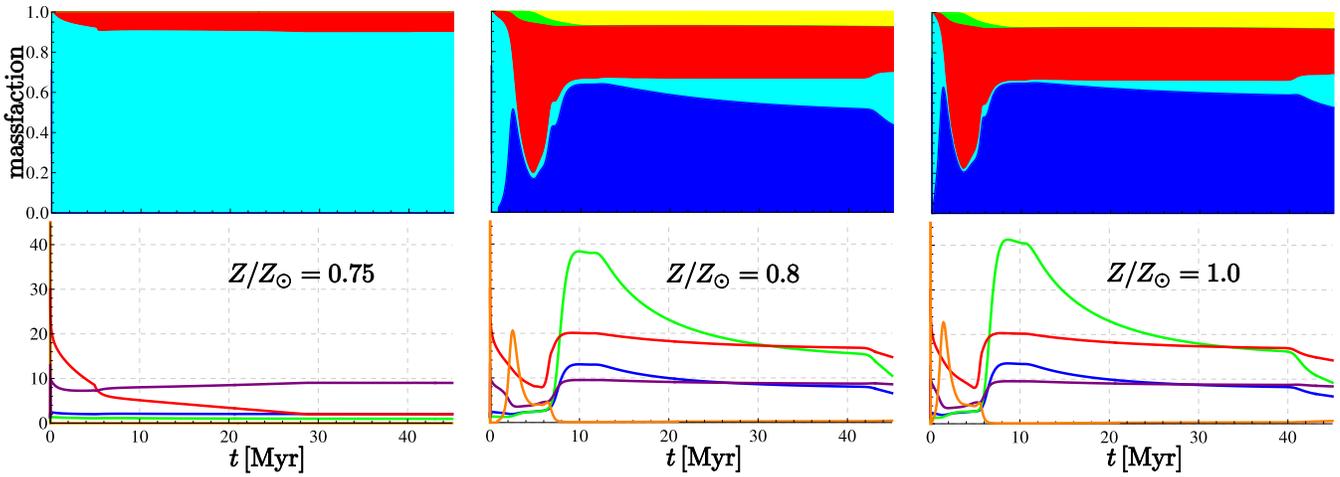}}
\caption{One-zone models for different metallicities at fixed initial density $n=75\,\mathrm{cm}^{-3}$ and moderate external driving $\mathcal{M}_\Sigma=0.2$. See the legend in Fig.~\ref{fig:SFtrunc_evol} for a definition of the plotted quantities.}
\label{fig:Z_evolution}
\end{minipage}\end{figure*}

\begin{figure*}\begin{minipage}{177mm}
  \resizebox{\hsize}{!}{\includegraphics{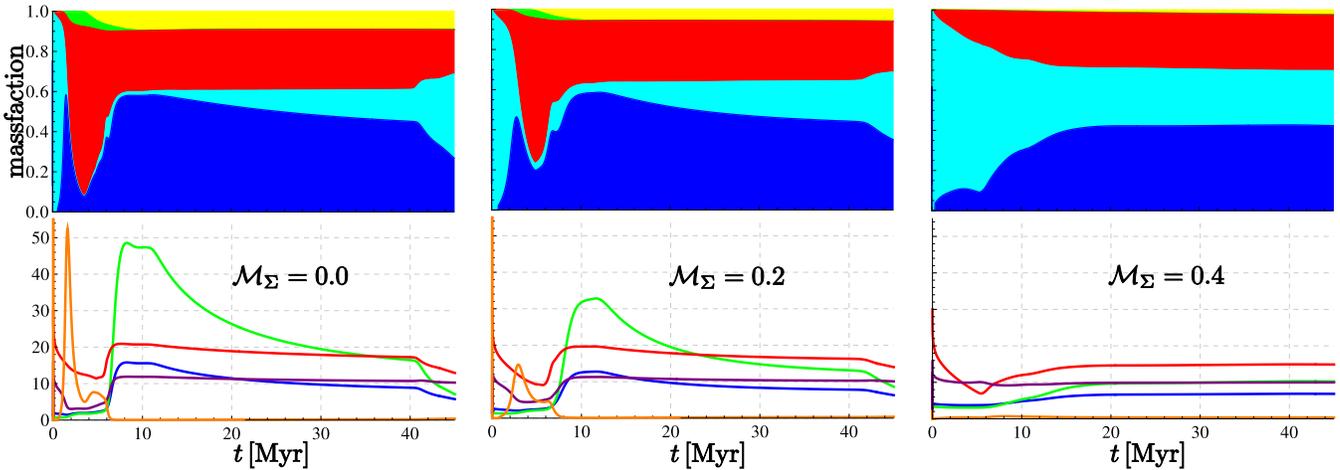}}
\caption{One-zone models for different rates of turbulent energy injection at fixed initial density $n=50\,\mathrm{cm}^{-3}$ and solar metallicity. See the legend in Fig.~\ref{fig:SFtrunc_evol} for a definition of the plotted quantities.}
\label{fig:M_evolution}
\end{minipage}\end{figure*}

By comparing Figs.~\ref{fig:ND_evolution} and~\ref{fig:Z_evolution}, where the latter figure shows plots for three one-zone models with an initial density $n=75\,\mathrm{cm}^{-3}$, but different metallicities $Z/Z_\odot=1.0,\,0.8,\,0.75$, one can see that lowering $Z$ has effects on the evolution similar to those of lowering $n$. This is simply a consequence of the dependence of the following processes on the density of metals, which is proportional to $nZ$:
\begin{itemize}
 \item $\Lambda_{\rm rad}$ is primarily determined by metal line cooling.
 \item The major contribution to the $\mathrm{H}_2$-production rate (see Eq.~\ref{eq:molfrac2}) is the formation of molecules on dust grains, whose abundance is assumed to be proportional to $Z$.
\item The absorption of $\mathrm{H}_2$-dissociating radiation outside of molecular cores is dominated by dust extinction.
\end{itemize}
Consequently, $f_{\rm c,H_2}$ and the relative mass fractions of phases are sensitive to $Z$, while quantities related to gravity are only indirectly affected. For this reason, $l_{\rm c}$, $t_{\rm ff}$ and $\SFR$ remain almost unaffected when varying $Z$.

Next, we consider the influence of turbulence driving. Generally, a higher production rate
$\Sigma$ increases the turbulent energy $e_{\rm t}$ and damps the peaks of star formation during the initial transient phase. Consequently, the stationary phase is entered earlier and more smoothly, as one can see in Fig.~\ref{fig:M_evolution}. Even in the absence of turbulent energy injection ($\mathcal{M}_\Sigma=0$), the turbulence generated by the cooling instability and SNe plays an important role in limiting the star formation rate (see the left panel in Fig.~\ref{fig:M_evolution}). This is caused by a decrease of $\rho_{\rm c,pa}$, as indicated by the growth of $l_{\rm c}$. Small $\mathcal{M}_{\rm \Sigma}$ do not cause a major increase of $e_{\rm t}$, $\mathcal{M}_{\rm c}$ or $l_{\rm c}$ (middle panel of Fig.~\ref{fig:M_evolution}) compared to $\mathcal{M}_\Sigma=0$, although $u_{\rm w}$ settles at a slightly higher level due to the increased energy input. However, if the production of turbulence is dominated by $\Sigma$, the turbulent contribution to the effective pressure becomes comparable to the thermal pressure in the warm phase ($e_{\rm t}\simeq u_{\rm w}$ in the right panel of Fig.~\ref{fig:M_evolution}). As a consequence, the cold-phase pressure increases relative to the warm-phase pressure (see Eq. ~\ref{eq:rho_c}), and $\rho_{c,{\rm pa}}$ decreases. This results a lower star formation efficiency $\varepsilon_{\rm ff}$. Apart from this effect on $\varepsilon_{\rm ff}$ for $e_{\rm t}\sim u_{\rm w}$, turbulence also affects the molecular fraction $f_{\rm c,H_2}$ and the star formation efficiency in the molecular gas, $\SFR$. These sub-dominant effects are discussed in Sect.~ \ref{sec:SFReq}.

\subsection{Equilibrium star formation efficiency}\label{sec:SFReq}

Although the transient phases discussed in Sect.~\ref{eq:dyn_evol} shed light on the complex interplay between the various physical processes contributing to the multi-phase dynamics, the approximately statistically stationary regimes of star formation are most relevant for applications. To numerically determine equilibrium solutions, we assume a small star formation rate of $1\ \%$ as initial condition and integrate the rate equations until the relative temporal variance of $\dot{\rho}_{\rm{s}}$ over a feedback period ($\gtrsim\!40$ Myr) becomes less than $10^{-4}$ or the star formation rate approaches zero (in the latter case the solution is not considered to be an equilibrium solution). We also check if the equilibrium conditions~(\ref{eq:rateuHeq}) and~(\ref{eq:rateeSGSeq}) are fulfilled with a relative accuracy better than $10^{-4}$. In contrast to the dynamical evolutions discussed before, we keep the total gas mass constant in the course of the integration, because we want to obtain equilibrium solutions for fixed gas densities that do not depend on the gas consumption by star formation during transient phases prior to the statistically stationary states. In numerical simulations, where the model describes a certain grid zone, one can think of the gas being replenished by neighbouring regions. Of course, such conditions will be met only to a certain degree and for a limited period of time. Nevertheless, the equilibrium solutions are useful to understand the behaviour of the system, and these solutions can be utilised as approximations to the star formation under quasi-stationary conditions.

\begin{figure*}
\centering
  \includegraphics[width=70mm]{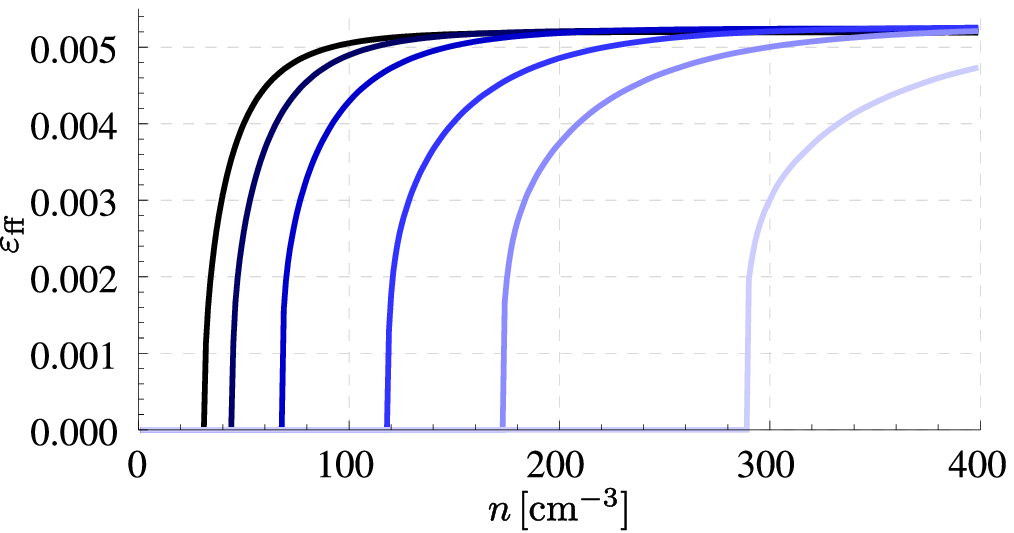}\hspace{18mm}
  \includegraphics[width=70mm]{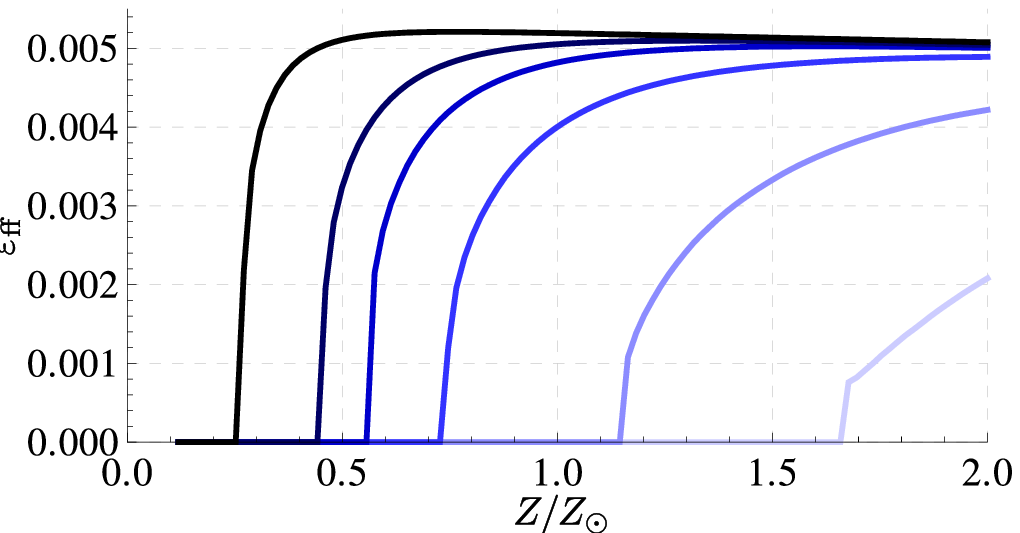}
\caption{Equilibrium star formation efficiency per free fall time, $\varepsilon_{\rm ff}$, versus number density $n$ for different metallicities $Z/Z_\odot= 1.0,\,0.8,\,0.6,\,0.4,\,0.2$ from dark to light blue (left panel), and $\varepsilon_{\rm ff}$ as function of $Z/Z_\odot$ for different $n=200,\,100,\,75,\,50,\,25,\,12.5\,\mathrm{cm}^{-3}$ from dark to light blue (right panel).}
\label{fig:epsFFvzZ_ND}
\end{figure*}

\subsubsection*{Gas density and metallicity}

Many star formation recipes used in astrophysical simulations assume, inspired by observational Kennicutt-Schmidt-relations, that stars are formed with a fixed efficiency per free fall time if a certain density threshold is exceeded. Our model shows such a behaviour under the condition of low turbulence driving $\Sigma$, as the star formation efficiency $\varepsilon_{\rm ff}$ saturates quickly above a metallicity-dependent density threshold (see Fig.~\ref{fig:epsFFvzZ_ND}, left panel). The dependence of $\varepsilon_{\rm ff}$ on $n$ is governed by the molecular fraction $f_{\rm c,H_2}$, while $\SFR$ remains approximately constant for all $n\gg1\,\mathrm{cm}^{-3}$. A similar dependence on the metallicity can be seen in Fig.~\ref{fig:epsFFvzZ_ND} right panel. This is a consequence of cooling, $\mathrm{H}_2$-production and extinction of $\mathrm{H}_2$-dissociating radiation being mainly dependent on $nZ$. 

For the zone evolutions with constant total mass (gas and stars) shown in the left panels of Figs.~\ref{fig:ND_evolution} and~\ref{fig:Z_evolution}, the star formation efficiency vanishes, although $n$ and $Z$ are above the threshold values for star formation following from Fig.~\ref{fig:epsFFvzZ_ND}. This can be understood as a consequence of the different initial conditions. Due to the lack of stellar feedback, the gas phases are never pushed into a star-forming regime in the former calculations. In a numerical simulation, the heating of gas by nearby stars could trigger star formation in so far inactive gas. Otherwise the gas will remain cold and inactive as long as it does not become dense enough to form molecular cores and to start star formation.

\begin{figure*}
\centering
  \includegraphics[width=70mm]{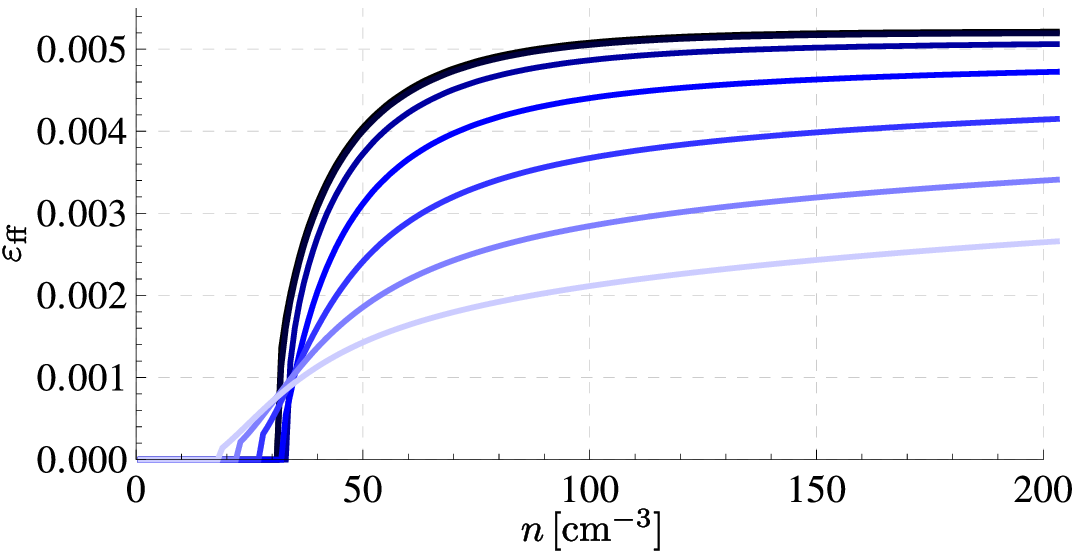}\hspace{18mm}
  \includegraphics[width=70mm]{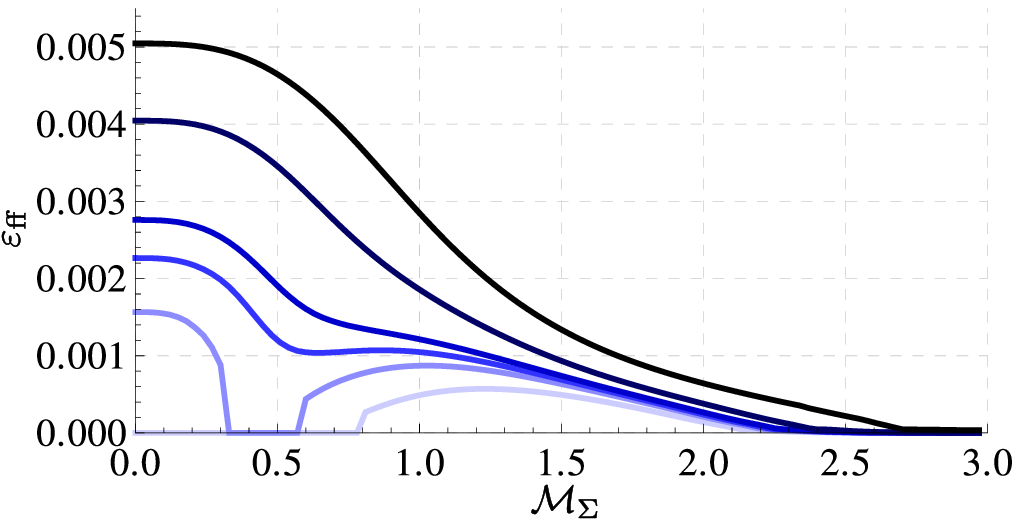}
\caption{Equilibrium star formation efficiency per free fall time, $\varepsilon_{\rm ff}$, versus number density $n$ for different turbulent energy injection rates $\Sigma$ (left panel) in terms of $\mathcal{M}_{\rm \Sigma}=0.0,\,0.2,\,0.4,\,0.6,\,0.8,\,1.0,\,1.2$ from dark to light blue (see Eq.~\ref{eq:mach_sigma}), and $\varepsilon_{\rm ff}$ as function of  $\mathcal{M}_{\rm \Sigma}$ for different number densities $n=27.25$, 32.5, 35, 37.5, 50, 100 $\mathrm{cm}^{-3}$ from light to dark blue (right panel).}
\label{fig:epsFFvzDriving_ND}
\end{figure*}

\subsubsection*{External turbulence driving}

Low turbulence intensity maintained by external driving ($\mathcal{M}_\Sigma\lesssim 0.2$) does not change the shape of $\varepsilon_{\rm ff}(n)$ appreciably. Fig.~\ref{fig:epsFFvzDriving_ND} (left panel) shows that moderate values of $\mathcal{M}_\Sigma$ increase the density threshold and lower the saturation level of $\varepsilon_{\rm ff}$ slightly. As $\mathcal{M}_\Sigma$ approaches unity, however, star formation becomes more and more inhibited. Turbulence also smears out the density threshold. In Fig.~\ref{fig:epsFFvzDriving_ND} (right panel), four regimes can be identified, in which the effects of turbulence injection differ with increasing $\mathcal{M}_\Sigma$:
\begin{enumerate}
\item For small $\mathcal{M}_\Sigma$, the additional heating of the gas by turbulent dissipation partially counters the effects of turbulence suppressing star formation that become dominant for stronger $\mathcal{M}_\Sigma$.

\item For higher $\mathcal{M}_\Sigma$, the production of $e_{\rm t}$ is dominated by the turbulent cascade and $e_{\rm t}$ roughly follows $\mathcal{M}_\Sigma^2$. As a consequence, turbulent pressure contributes significantly to the pressure balance between the phases, and $\rho_{\rm c,pa}$ decreases ($\rho_{\rm c,pa}\rightarrow\rho$ in the limit of large $e_{\rm t}$). The lowering of the cold-gas density results in the steep reduction of $\varepsilon_{\rm ff}$ as $\mathcal{M}_\Sigma$ rises.

\item For stronger turbulence intensity, $\rho_{\rm c,pa}$ is low but the turbulent broadening of the density pdf becomes important (enhancement of $\mathrm{H_2}$-production by the clumping factor $C_{\rm \rho}$ and dependence of $\SFR$ on the pdf; see Sect.~\ref{sec:SFR}). This effect can clearly be seen for initial densities lower than $50\,\mathrm{cm}^{-3}$, where a second maximum of $\varepsilon_{\rm ff}$ can be discerned. This is the regime, in which star formation critically depends on the properties of self-gravitating turbulence.
 
\item If $\mathcal{M}_\Sigma$ increases further, the growth of of the minimum overdensity for star formation in the cold gas, $x_{\rm crit}\propto e_{\rm t}^2$ (see Eqns.~\ref{eq:x_crit} and~\ref{eq:sfr}) dominates, and $\varepsilon_{\rm ff}$ asymptotically falls off to zero.

\end{enumerate}

\begin{figure*}
\centering
  \includegraphics[width=70mm]{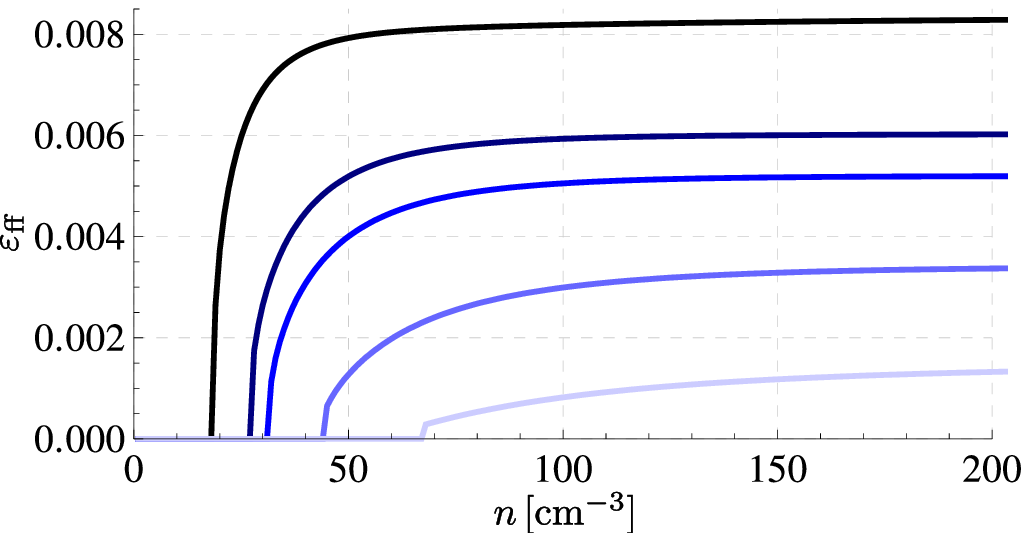}\hspace{18mm}
  \includegraphics[width=70mm]{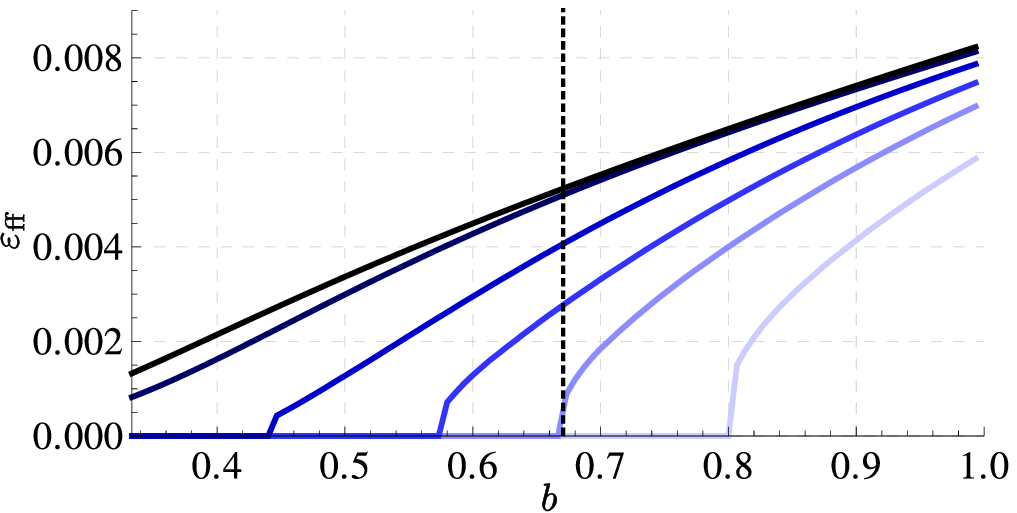}
\caption{Equilibrium star formation efficiency per free fall time, $\varepsilon_{\rm ff}$, versus number density $n$ for different values of the compressive factor $b$ in the cold phase (left panel). From light to dark blue, $b=1/3,\,0.5,\,2/3,\,0.75,\,1.0$. $\varepsilon_{\rm ff}$ over turbulence forcing parameter $b$ (right panel)  for the number densities $n=200,\,100,\,50,\,37.5,\,31.25,\,25\,\mathrm{cm}^{-3}$ (from dark to light blue). The default value $b=2/3$ is marked by the dotted line.}
\label{fig:epsFFvzND_bCompressive}
\end{figure*}

\begin{figure*}
\centering
\includegraphics[width=70mm]{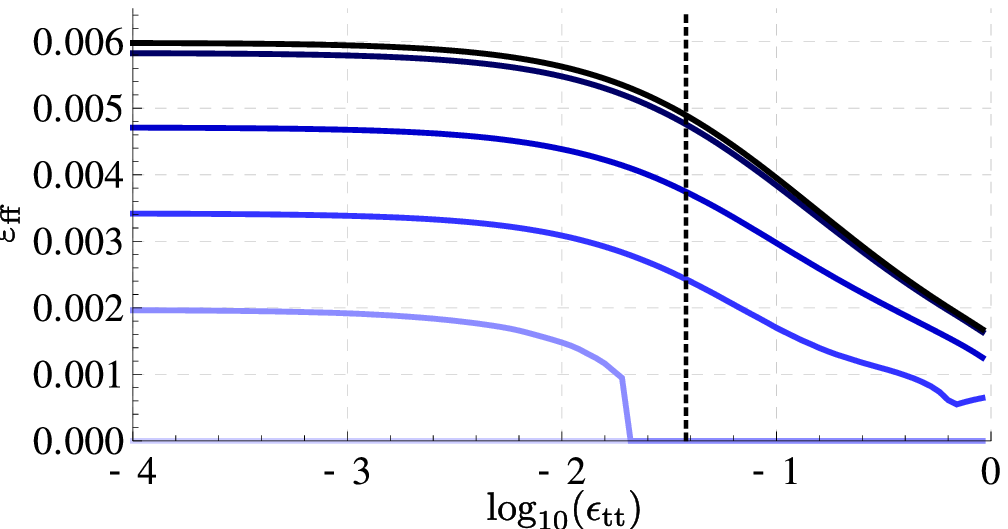}\hspace{18mm}
\includegraphics[width=70mm]{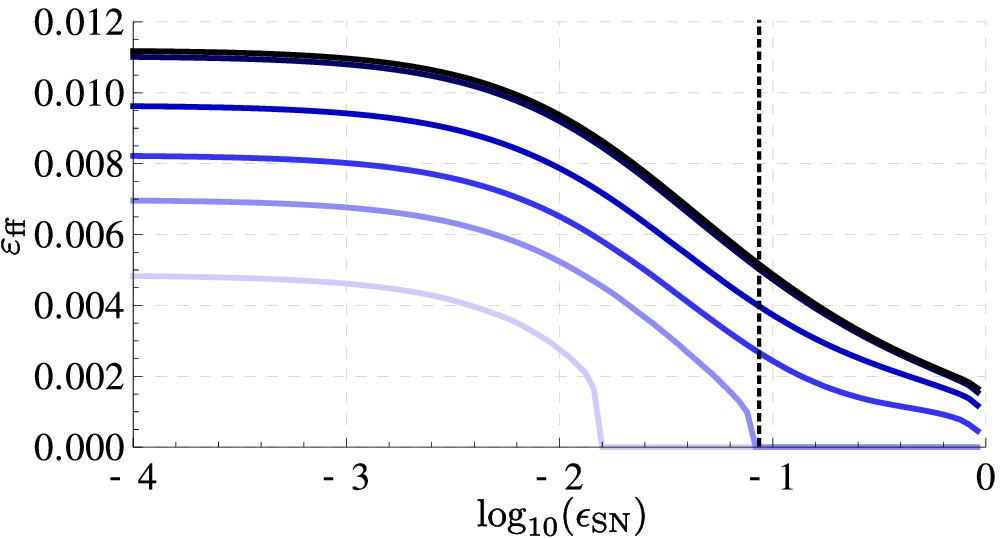}

\caption{Equilibrium star formation efficiency per free fall time, $\varepsilon_{\rm ff}$, versus the turbulent production efficiencies $\epsilon_{\rm tt}$ (left panel), and $\epsilon_{\rm SN}$ (right panel) for the number densities $n=200,\,100,\,50,\,37.5,\,31.25,\,25\,\mathrm{cm}^{-3}$ (from dark to light blue). The default values ($\epsilon_{\rm tt}=0.025$, $\epsilon_{\rm SN}=0.085$) are marked by the dotted lines.}
\label{fig:EvsTTSNb_N}
\end{figure*}

Because of the dependence of $\SFR$ and $f_{\rm c,H_2}$ on the density pdf, the saturation level of $\varepsilon_{\rm ff}$ and the density threshold are significantly affected by the weight of compressive forcing modes relative to solenoidal modes. Assuming that the width of the density pdf is given by $\sigma\approx\log(1+\mathcal{M}_{\rm c}^2b^2)$, where $\mathcal{M}_{\rm c}$ is the the rms-Mach number of turbulent motions in the cold gas and $b$ varies between $1/3$ for purely solenoidal forcing and $1$ for purely compressive forcing, we obtain the equilibrium solutions plotted in Figs.~\ref{fig:epsFFvzND_bCompressive} and~\ref{fig:EvsTTSNb_N} (right panel). The nature of turbulence driving in the interstellar medium is still a matter of debate. Moreover, the mixture of solenoidal and compressive modes is likely to be scale-dependent. Here, we adopt the intermediate value $b=2/3$ as default.

\begin{figure}
\centering
  \resizebox{0.8\hsize}{!}{\includegraphics{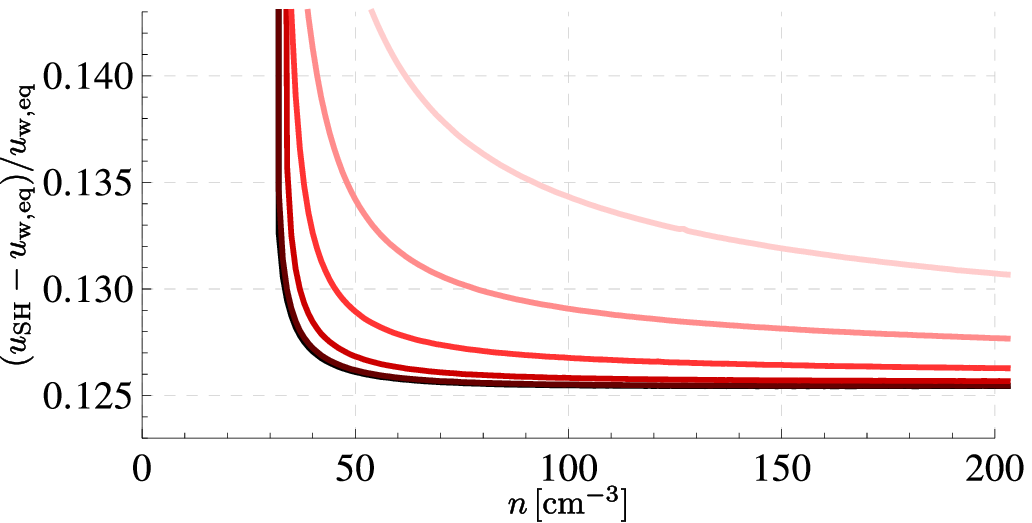}}
\caption{Relative deviation of the equilibrium solution for the warm-gas energy, $u_{\rm w,eq}$, from the first two terms on the right-hand side of Eq.~(\ref{eq:uw_eq}), corresponding to zero turbulent energy. The different curves are obtained for $\mathcal{M}_{\rm \Sigma}=$ 0.0, 0.2, 0.4, 0.6, 0.8, 1.0 from dark to light red.}
\label{fig:uDiffVSn_Sigma}
\end{figure}

As explained in Sect.~\ref{sc:equilbr}, turbulence generally decreases the temperature of the warm gas in equilibrium (see Eq.~\ref{eq:uw_eq}). Fig.~\ref{fig:uDiffVSn_Sigma} shows that this is roughly a $10\,\%$ effect for densities higher than the threshold for star formation. In a certain sense, this is the deviation from the SH03 equilibrium solution $u_{\rm SH}$ (see Eq.~\ref{eq:u_SH}). However, the value of $u_{\rm{SN}}$ used by SH03 and the coefficient of the first term is different from our definition so that the actual difference is greater. As one can see in Fig.~\ref{fig:uDiffVSn_Sigma}, there is a noticeable deviation even without external turbulence driving. This effect is caused by internal turbulence driving, which is considered next.

\begin{figure}
\centering
  \resizebox{0.8\hsize}{!}{\includegraphics{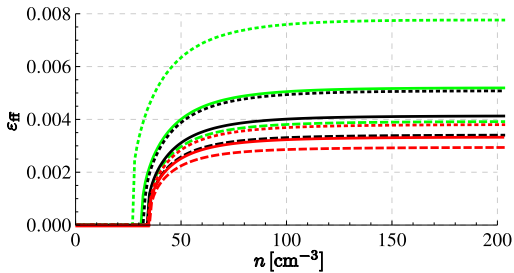}}
\caption{Equilibrium star formation efficiency per free fall time, $\varepsilon_{\rm ff}$, versus number density $n$ for different values of the turbulence production efficiencies $\epsilon_{\rm tt}$ and $\epsilon_{\rm SN}$: $\epsilon_{\rm tt}=0.025$ (green), $0.085$ (black), $0.17$ (red), and $\epsilon_{\rm SN}=0.025$ (dotted), $0.085$ (solid), $0.17$ (dashed).}
\label{fig:epsFFvzND_epsTTepsSN}
\end{figure}

\subsubsection*{Thermal instability and stellar feedback}

If the energy transfer from larger scales is small, internal driving by the thermal instability and stellar feedback dominate the production of turbulence. In this case, $e_{\rm t}$ is mainly controlled by the turbulence production efficiencies $\epsilon_{\rm tt}$ and $\epsilon_{\rm SN}$. An increase of $\epsilon_{\rm tt}$ lowers the thermal energy of the warm gas, $u_{\rm w}$, as the cooling instability transfers thermal energy to turbulence more efficiently. For higher $\epsilon_{\rm SN}$, less energy is deposited by SNe in the warm phase and also more turbulent energy is produced, which tends to decrease the star formation rate. However, the effect of $\epsilon_{\rm SN}$ is limited for high $\epsilon_{\rm tt}$ because the increased production of turbulent energy by the thermal instability reduces star formation and, consequently, supernova feedback. On the other hand, only a relatively small fraction of the thermal energy of the warm phase can be converted into turbulent energy by the thermal instability without violating the second law of thermodynamics. This suggests that $\epsilon_{\rm tt}$ has to be small compared to unity. 

Since turbulence produced by SNe on length scales much smaller than $l$ is rapidly dissipated into thermal energy, only the fraction specified by $\epsilon_{\rm SN}$ effectively enters the turbulent energy $e_{\rm t}$, while the fraction $1-\epsilon_{\rm SN}$ is immediately turned into thermal energy. This implies that $\epsilon_{\rm SN}$ is scale-dependent. For $l$ greater than a few parsec, $\epsilon_{\rm SN}$ must not be lower than a few percent. Otherwise the star formation efficiency would become significantly greater than $0.1$, in contradiction to the majority of observations \citep{KrumTan07,Murray10}. The average total energy deposited by a supernova in the interstellar medium is $E_{\rm SN}\approx10^{51}\,\mathrm{erg}$. Roughly $8.5\cdot10^{49}\;\mathrm{erg}$ of this energy enter the ISM in form of kinetic energy, or from the perspective of our model in form of turbulent energy \citep{Thornton98}. Hence, we adopt the value $\epsilon_{\rm SN}=0.085$ as default. The left and middle panels of Fig.~\ref{fig:EvsTTSNb_N} show that $\varepsilon_{\rm ff}$ saturates if one efficiency is much greater than the other. This is a consequence of $\rho_{\rm c,pa}\!\rightarrow\!\rho$, if turbulent energy is efficiently produced by whatever mechanism. To obtain a plausible star formation efficiency, we choose $\epsilon_{\rm tt}= 0.025$.

\begin{figure}
\centering
  \resizebox{0.8\hsize}{!}{\includegraphics{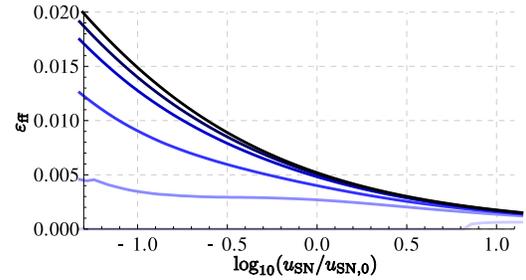}}
\caption{Equilibrium star formation efficiency per free fall time, $\varepsilon_{\rm ff}$, versus logarithmic specific energy of SN $\log(u_{\rm SN}/u_{\rm SN,0})$, where $u_{\rm SN,0}=6\!\cdot\!10^{49}\mathrm{erg}/M_\odot$, for $n=$ 25, 37.5, 50, 75, 100, 200 $\mathrm{cm}^{-3}$ from light to dark blue.}
\label{fig:epsFFvzuSN_N}
\end{figure}

With the number of SNe per solar mass of formed stars,
\begin{equation}
 n_{\rm SN}=\frac{1}{M_*}\int_{8M_\odot}^{40M_\odot}\frac{\mathrm{d}N}{\mathrm{d}m_*}\mathrm{d}m_*,
\end{equation}

we estimate the feedback energy per solar mass:
\begin{equation}
 u_{\rm SN}=n_{\rm SN}E_{\rm SN}\frac{1-\beta}{\beta}\approx6\!\cdot\!10^{49}\,\mathrm{erg}/M_{\odot},
\end{equation}
with the feedback fraction $\beta$ as defined in Eq.~\ref{eq:beta}. This value is subject to large uncertainties. However, Fig.~\ref{fig:epsFFvzuSN_N} shows that the star formation efficiency is relatively robust against variations in $u_{\rm SN}$ if its value is at least the same order of magnitude as the above default value for the model.

\begin{figure}
\centering
  \resizebox{0.8\hsize}{!}{\includegraphics{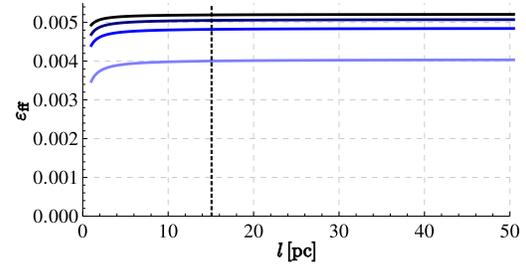}}
\caption{Equilibrium star formation efficiency per free fall time, $\varepsilon_{\rm ff}$, versus length scale $l$ for different number densities $n=50,\,75,\,100,\,200\,\mathrm{cm}^{-3}$ (from light to dark blue), without external driving.}
\label{fig:epsFFvzND_l}
\end{figure}

\begin{figure}
\centering
  \resizebox{0.8\hsize}{!}{\includegraphics{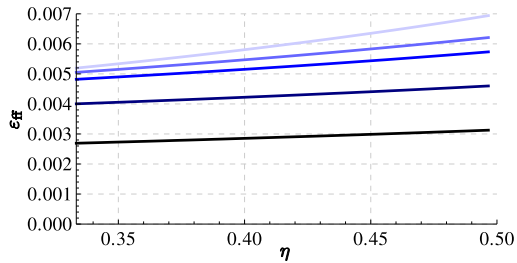}}
\caption{Equilibrium star formation efficiency per free fall time, $\varepsilon_{\rm ff}$, versus the turbulence scaling exponent $\eta$ for $n=37.5,\,50,\,75,\,100,\,200\,\mathrm{cm}^{-3}$ from dark to light blue.}
\label{fig:epsFFvzEta_N}
\end{figure}

\subsubsection*{Scale dependence}

For length scales $l\gtrsim10\,\mathrm{pc}$, the star formation efficiency is almost exactly scale-invariant (see Fig.~\ref{fig:epsFFvzND_l}). This demonstrates that even without maintaining a scale-invariant rate of energy injection by external turbulence forcing, the system settles into an equilibrium state, for which the clump length scale $l_{\rm c}$ and the Mach number of turbulence in the cold-gas phase, $\mathcal{M}_{\rm c}$, are nearly independent of $l$. Internal driving and turbulent dissipation regulate $e_{\rm t}$ such that the scaling of the turbulent velocity fluctuations from $l$ to $l_{\rm c}$ by the power law~(\ref{eq:sigma_c}) results in a fixed $\mathcal{M}_{\rm c}\propto\sigma_{\rm c}$, regardless of the choice of $l$. However, the equilibrium is influenced by the scaling parameter $\eta$. For weakly compressible turbulence (up to Mach numbers around unity), $\eta$ is close to the Kolmogorov value $1/3$, whereas $\eta$ rises to $1/2$ for supersonic turbulence. The dependence of $\varepsilon_{\rm ff}$ on $\eta$ is shown in Fig.~\ref{fig:epsFFvzEta_N}. Since we assume that turbulence is supersonic within the cold clumps, but transonic in the warm gas, and there are only little changes of $\varepsilon_{\rm ff}$ if $\eta$ is about $1/3$, we set $\eta=1/3$ as a reasonable approximation if $l$ is greater than $l_{\rm c}$.

If $l$ is only a few parsec or less, on the other hand, $l_{\rm c}$ may exceed $l$. In this case, basic assumptions of the model break down, as the notion of cold-gas clumps in pressure balance with the warm gas in the reference volume of size $l$ becomes meaningless. The percolation of the cold phase and the transition to a one-phase medium is not yet implemented in the model (see Sect.~\ref{sec:onephase}). If the model in its present form is applied as an SGS model in AMR simulations, a maximum refinement limit has to be applied such that $l_{\rm c}<l$ is ensured. In cosmological simulations with $l_{\rm min}\sim 100\,\mathrm{pc}$, this condition will almost certainly be satisfied. For high-resolution simulations of individual galaxies with $l_{\rm min}\sim 1\,\mathrm{pc}$, however, situations, where cold-gas regions extend over several zones cannot be avoided. Then a viable model has to deal the transition to a one-phase medium. 

\begin{figure*}
\centering
  \includegraphics[width=70mm]{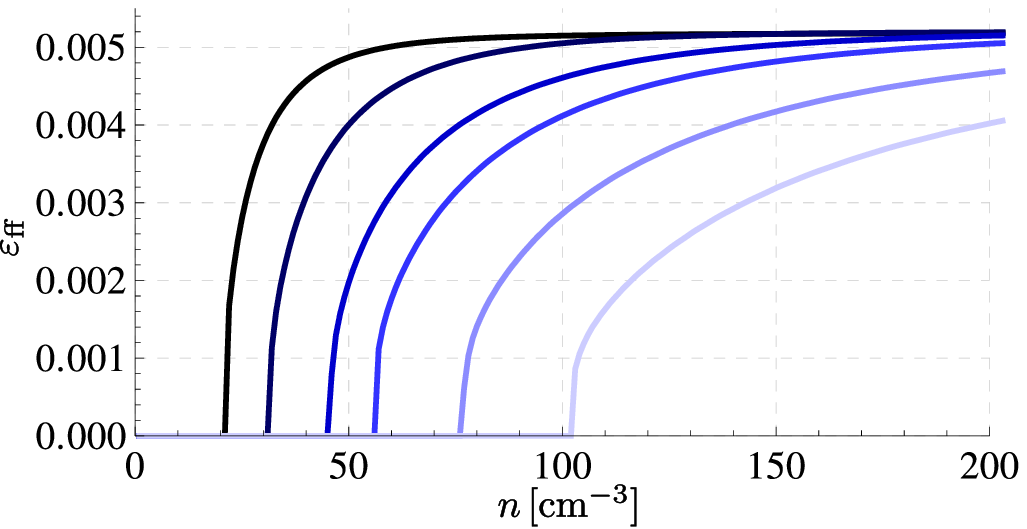}\hspace{18mm}
  \includegraphics[width=70mm]{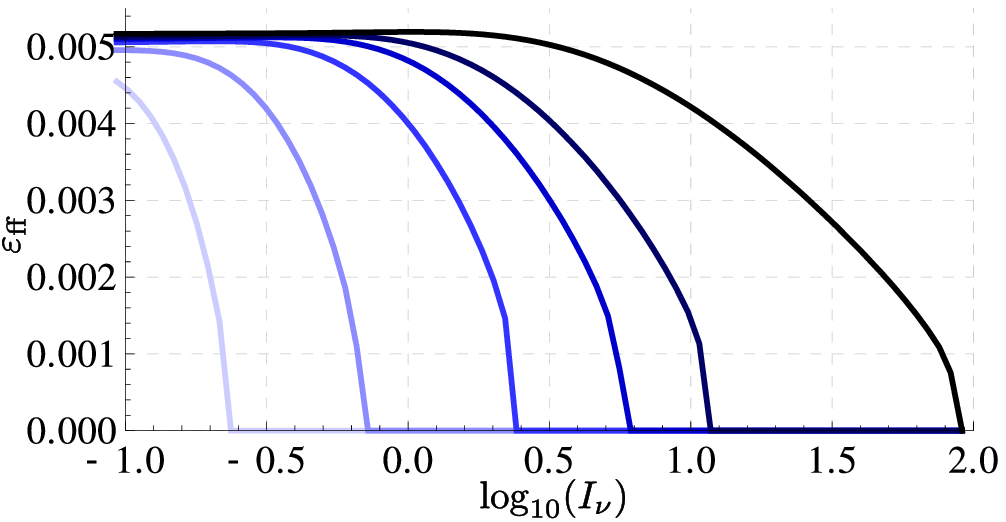}
\caption{Equilibrium star formation efficiency per free fall time, $\varepsilon_{\rm ff}$, versus number density $n$ for the normalised interstellar radiation field $I_\nu=0.5,\,1.0,\,1.5,\,2.0,\,3.0$ from dark to light blue (left panel). The metallicity of the gas is $Z=0.5\!\cdot\!Z_\odot$, and $\varepsilon_{\rm ff}$ as function of $\log(I_\nu)$ for $n=\,25,\,50,\,75,\,100,\,200\,\mathrm{cm}^{-3}$ from light to dark blue (right panel). The default $I_\nu=1$ is marked by the vertical dotted line. 
%[\textbf{Axis label is wrong! no, it isn't}]
}
\label{fig:epsFFvzDissRad_ND}
\end{figure*}

\subsubsection*{Photo-dissociation and heating}

The intensity of the ambient interstellar radiation field $I_{\nu}$ (in units of the Draine-field) is an external parameter and depends on the environment of the reference volume, i.~e., the location and structure of the host galaxy and its surroundings. Thus, it is a parameter at the same level as $n$, $Z$ and $\mathcal{M}_{\rm \Sigma}$, which has a direct impact on $f_{\rm c,H_2}$ and $u_{\rm w}$ via $\mathrm{H_2}$-dissociation and photoelectric heating. Fig.~\ref{fig:epsFFvzDissRad_ND} demonstrate that the density threshold of star formation and the shape of $\varepsilon_{\rm ff}(n)$ change significantly with variations in $I_\nu$, while the saturation level of $\varepsilon_{\rm ff}$ remains nearly unaffected. Consequently, the threshold density of star formation is mainly determined by $Z$ and $I_\nu$.

\begin{figure*}
\centering
  \includegraphics[width=70mm]{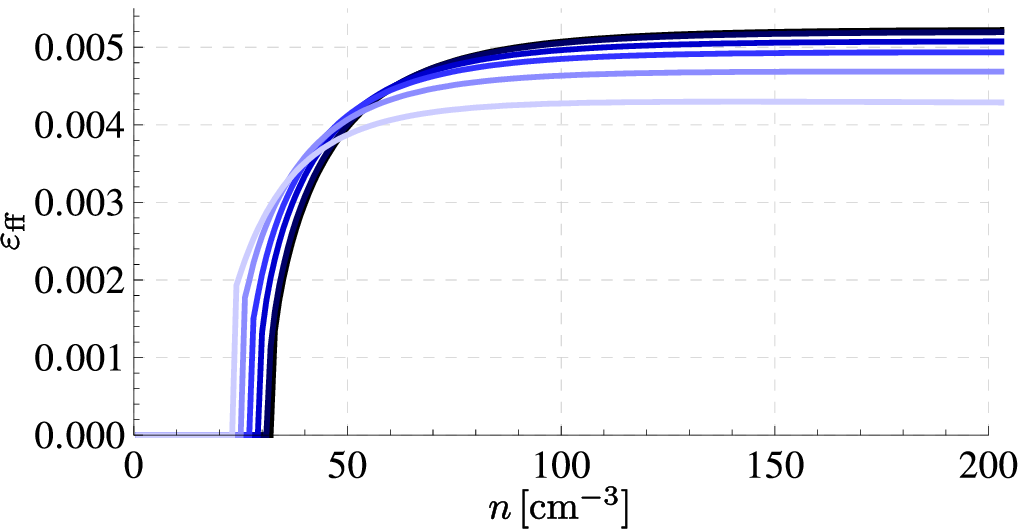} \hspace{18mm}
  \includegraphics[width=70mm]{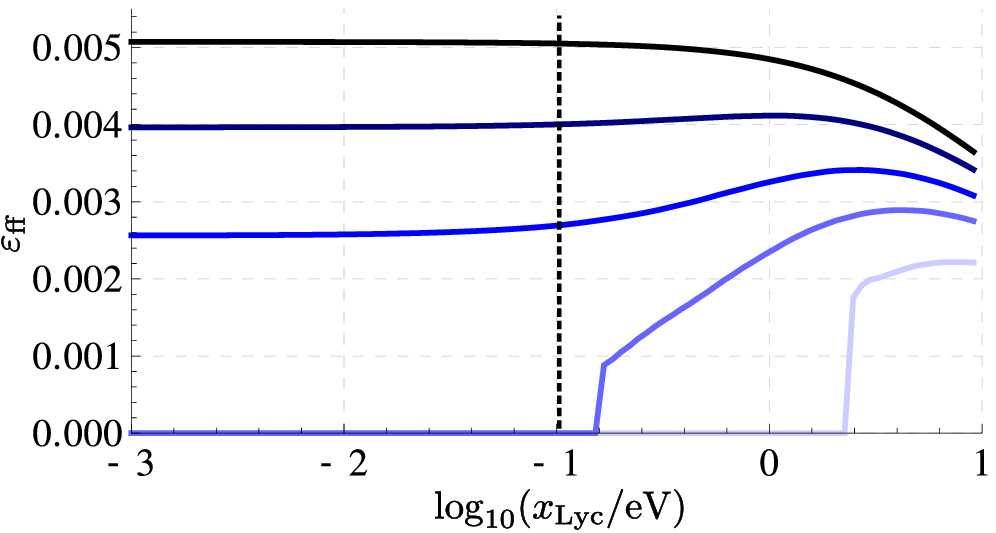}
\caption{Equilibrium star formation efficiency per free fall time, $\varepsilon_{\rm ff}$, versus number density $n$ for $x_{\rm Lyc}=$ 0.01, 0.1, 0.5, 1.0, 2.0, 4.0 $\mbox{ eV/photon}$ from dark to light blue (left panel), and $\varepsilon_{\rm ff}$ as function of $\log(x_{\rm Lyc}/\mathrm{eV})$ for different  $n=25$, 26.5, 28.125, 31.25, 37.5, 50, 100 $\mathrm{cm}^{-3}$ from light to dark blue (right panel). The default $x_{\rm Lyc}=0.1\,\mathrm{eV}$ is marked by the dotted line.}
\label{fig:epsFFvzxLyc_N}
\end{figure*}

The heat deposited in the gas per Lyc-photon, $x_{\rm Lyc}$, depends on the shape of the spectrum emitted by the young massive stars and the state of the absorbing gas. Both gas phases are affected by Lyc-heating (see Sect.~\ref{sec:evolrho}), and it reduces the amount of warm gas dropping into the cold phase if the cooling instability is active. As a consequence, an increase of $x_{\rm Lyc}$ lowers the level of saturation of $\varepsilon_{\rm ff}(n)$ significantly (see Fig.~\ref{fig:epsFFvzxLyc_N}). Even for small $x_{\rm Lyc}$, the density threshold of star formation is affected by the influence of Lyc-heating on the thermal pressure of the warm phase and, thus, on $\rho_{\rm c,pa}$. Since the cross section decreases significantly for large excess energies above the Rydberg energy and inelastic scattering processes become more likely, $x_{\rm Lyc}>0.5\,\mathrm{eV}$ is not plausible. We use $x_{\rm Lyc}=0.1\,\mathrm{eV}$ as default value.

\subsubsection*{Clump collisions}

The evaporation of clumps due to collisions lowers the $\rho_{\rm c}/\rho_{\rm w}$ ratio, but the clumps are typically too small and collisions are too rare to influence the evolution significantly even for a high efficiency $\epsilon_{\rm cc}$. If $l_{\rm c}$ becomes comparable to $l$, clump collisions cannot be applied for obvious reasons. We thus neglect this effect altogether by setting $\epsilon_{\rm cc}=0$.

\subsubsection*{Prestellar mass loss}

\begin{figure*}
\centering
  \includegraphics[width=70mm]{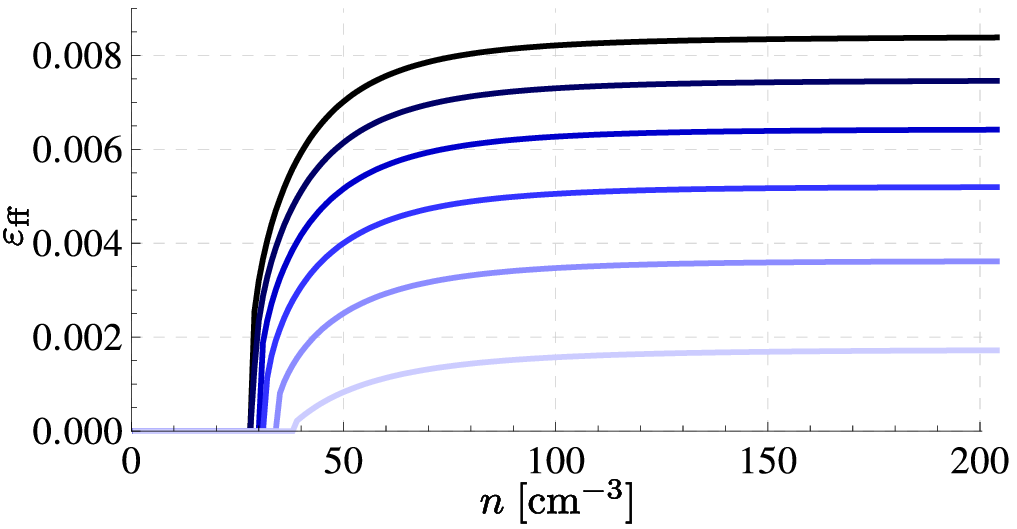}
\hspace{18mm}
  \includegraphics[width=70mm]{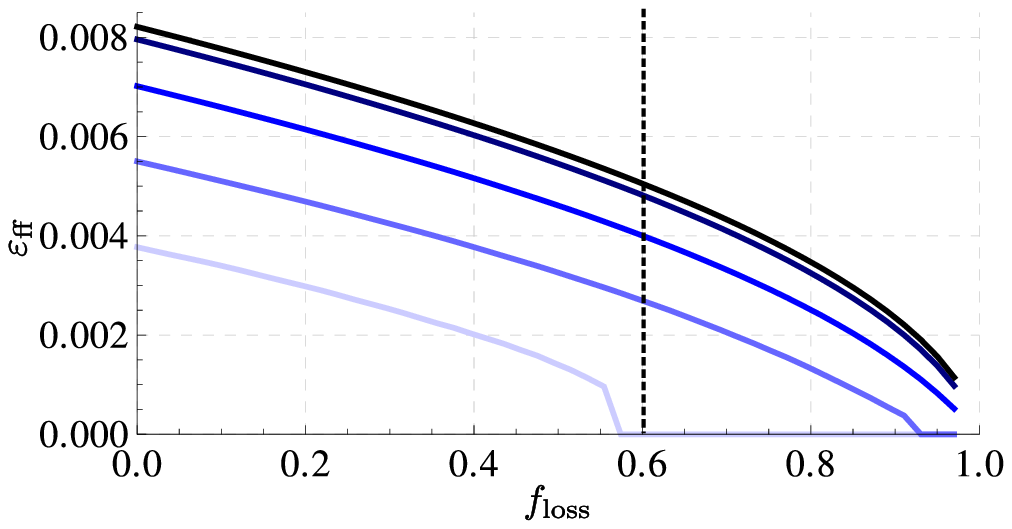}
\caption{Equilibrium star formation efficiency per free fall time, $\varepsilon_{\rm ff}$, versus number density $n$ for $f_{\rm loss}=$ 0.0, 0.2, 0.4, 0.6, 0.8, 0.95  from dark to light blue (left panel), and $\varepsilon_{\rm ff}$ as function of $f_{\rm loss}$ for different  $n=31.25,\,37.5,\,50,\,75,\,100\,\mathrm{cm}^{-3}$ from light to dark blue (right panel). The default $f_{\rm loss}=0.5$ is marked by the dotted line.}
\label{fig:epsFFvzfLoss_N}
\end{figure*}

As mentioned by \citet{Chabrier2010}, the CMF (the observed mass function of gravitationally bound cores in molecular clouds) and the IMF (the initial mass function of stars) are similar, except for an almost mass-independent shift by factor about $2-3$ \citep[also see][]{Matzner2000}. We account for the mass reduction due to the evolution from the CMF to the IMF by reducing the star formation efficiency by a factor $1-f_{\rm loss}$, where $f_{\rm loss}$ is interpreted as the mass fraction that is ejected during the collapse prior to star formation. As one can see in Fig.~\ref{fig:epsFFvzfLoss_N}, increasing $f_{\rm loss}$ reduces the star formation rate significantly. A good agreement with observational relations is obtained for the intermediate value $f_{\rm loss}=0.6$.

\subsection{Comparison to observations}
\label{sec:comparison}
Comparisons of one-zone results with observations are difficult, mainly for the following reasons. Firstly, the conversion of the modelled volume densities into the corresponding surface densities is nontrivial. Secondly, star forming regions are generally not in local star formation equilibrium, which could explain the breakdown of Kennicutt-Schmidt relations on small scales \citep[e.g.][]{Onodera2010,Schruba2010,Murray10}. Without the detailed dynamic environment in a hydrodynamic simulation, we can only draw conclusions on the basis of the equilibrium solutions calculated with our model. Even so, we are able to demonstrate that these solutions are consistent with the constraints set by observations on kpc scales.

\begin{figure}
\centering
  \resizebox{0.8\hsize}{!}{\includegraphics{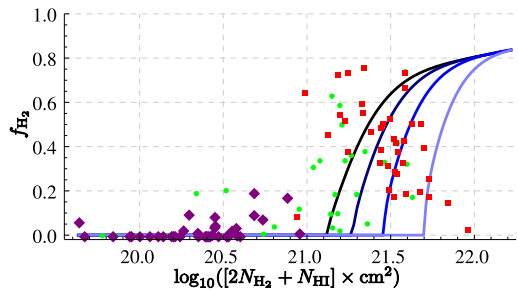}}
\caption{Molecular fraction $f_{\rm H_2,tot}$ versus logarithmic total hydrogen surface number density $\log(2N_{\rm H_2}+N_{\rm HI})$ in star formation equilibrium for different $Z/Z_\odot=$ 1.0, 0.8, 0.6, 0.4 from dark to light blue. Red squares \citep{Rachford2002,Rachford2009}, green circles \citep{Wolfire2008}, and purple diamonds \citep{Gillmon2006} represent observations of translucent clouds in the Galaxy.}
\label{fig:fH2vsN_obs}
\end{figure}

Since the calculation of $f_{\rm c,H_2}$ in our model only treats molecular hydrogen in cold clumps shielded from UV-radiation, we neglect the molecular hydrogen in radiation-dominated areas, where a certain amount exists in equilibrium between production and radiative destruction. As a consequence, the predicted molecular hydrogen mass might be systematically too low. This discrepancy can become particularly strong in the case of low (column) densities. To estimate the hydrogen column density, we set $(2N_{\rm H_2}+N_{\rm HI})\simeq X\rho l/m_{\rm H}$. Fig.~\ref{fig:fH2vsN_obs} shows that the transition from marginal to significant total molecular fractions $f_{\rm H_2,tot}$ occurs at column densities that are in good agreement with observations of translucent clouds in the Milky Way.

\begin{figure*}
\centering
\includegraphics[width=70mm]{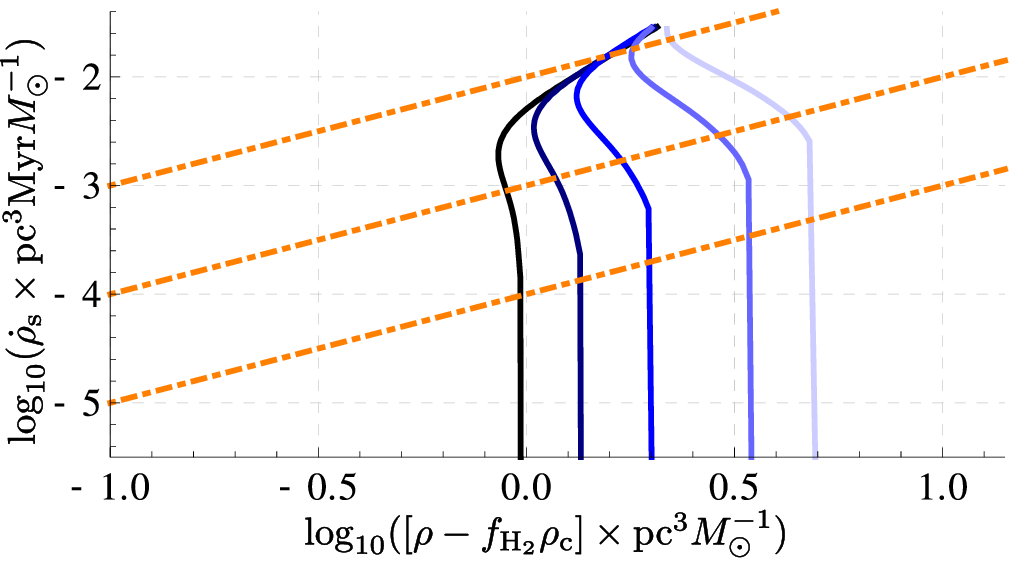}\hspace{18mm}
\includegraphics[width=70mm]{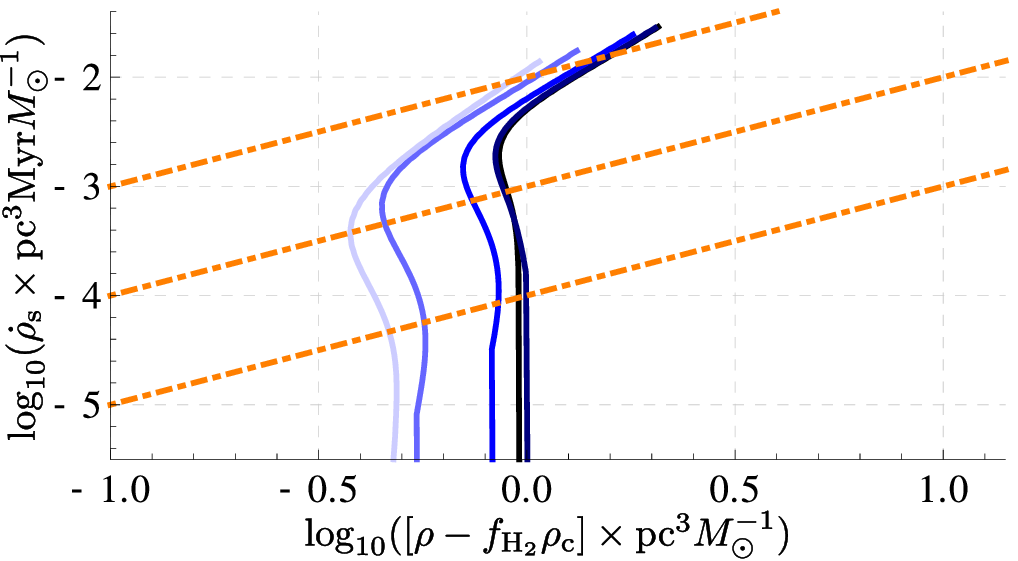}
\caption{Star formation rate as function of the fractional atomic density $\rho-f_{\rm c,H_2}\rho_{\rm c}$ for $Z/Z_\odot= 1.0,\,0.8,\,0.6,\,0.4,\,0.3$ from dark to light blue (left panel) and for different turbulent energy injection rates $\Sigma$ in terms of $\mathcal{M}_{\rm \Sigma}=0.0,\,0.4,\,0.8,\,1.2,\,1.4$ (from dark to light blue, right panel). The long- and short-dashed lines indicate the asymptotes with slopes $1.4$ ($\mathcal{M}_{\rm \Sigma}=0.0$, all $Z/Z_\odot$) and $1.6$ ($\mathcal{M}_{\rm \Sigma}=1.0$), respectively.  The orange dot-dashed lines mark fixed depletion time scales of 0.1, 1, and 10 Gyr from the top to the bottom of the graph.}
\label{fig:rhoSdotVSrhoHI}
\end{figure*}

\begin{figure*}
\centering
\includegraphics[width=70mm]{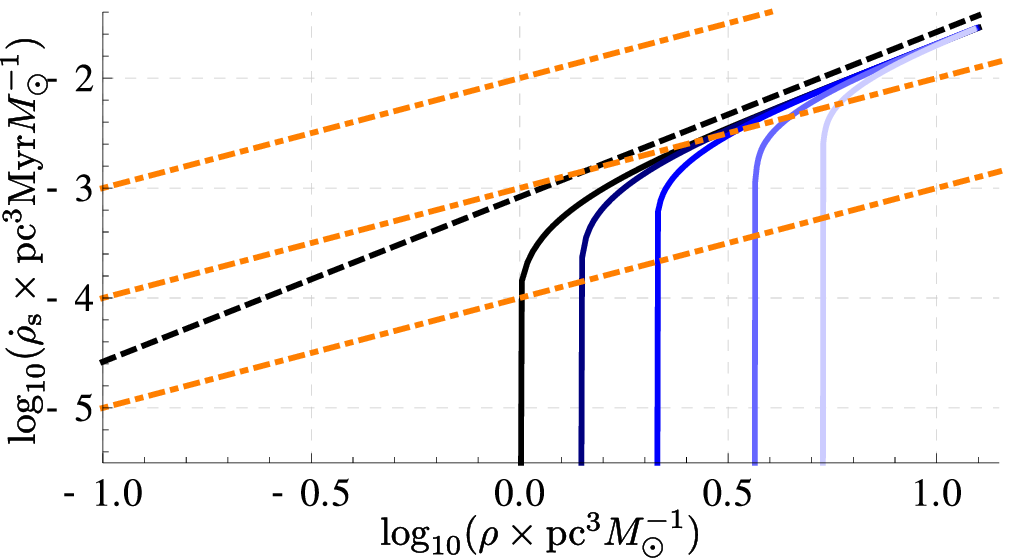}\hspace{18mm}
\includegraphics[width=70mm]{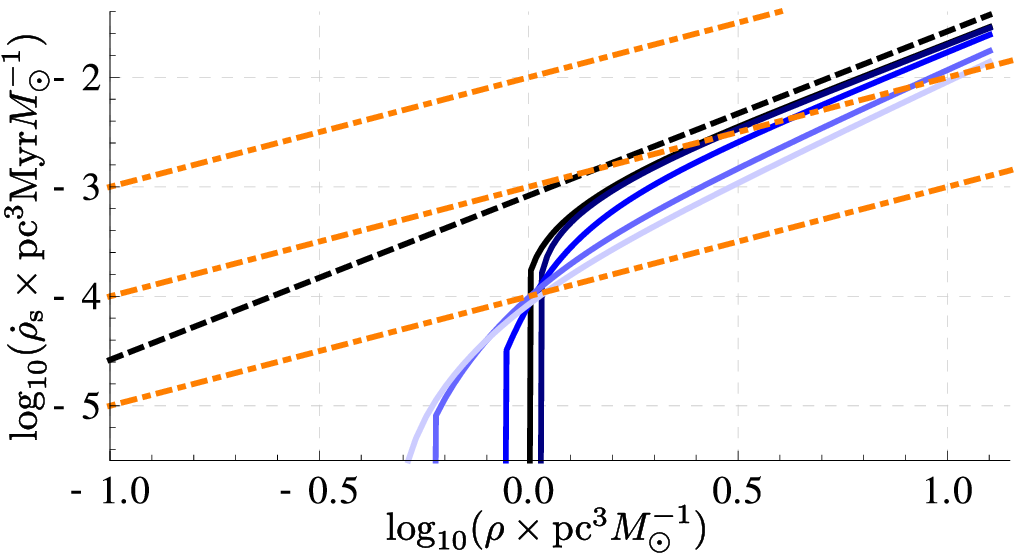}
\caption{Star formation rate as function of the total density $\rho$ for $Z/Z_\odot= 1.0,\,0.8,\,0.6,\,0.4,\,0.3$ from dark to light blue (left panel) and for different turbulent energy injection rates $\Sigma$ in terms of $\mathcal{M}_{\rm \Sigma}=0.0,\,0.4,\,0.8,\,1.2,\,1.4$ (from dark to light blue, right panel). The Kennicutt-Schmidt-asymptote with slope $1.5$ is indicated by the dashed line, orange dot-dashed lines as in Fig.~\ref{fig:rhoSdotVSrhoHI}.}
\label{fig:rhoSdotVSrho}
\end{figure*}

\begin{figure*}
\centering
\includegraphics[width=70mm]{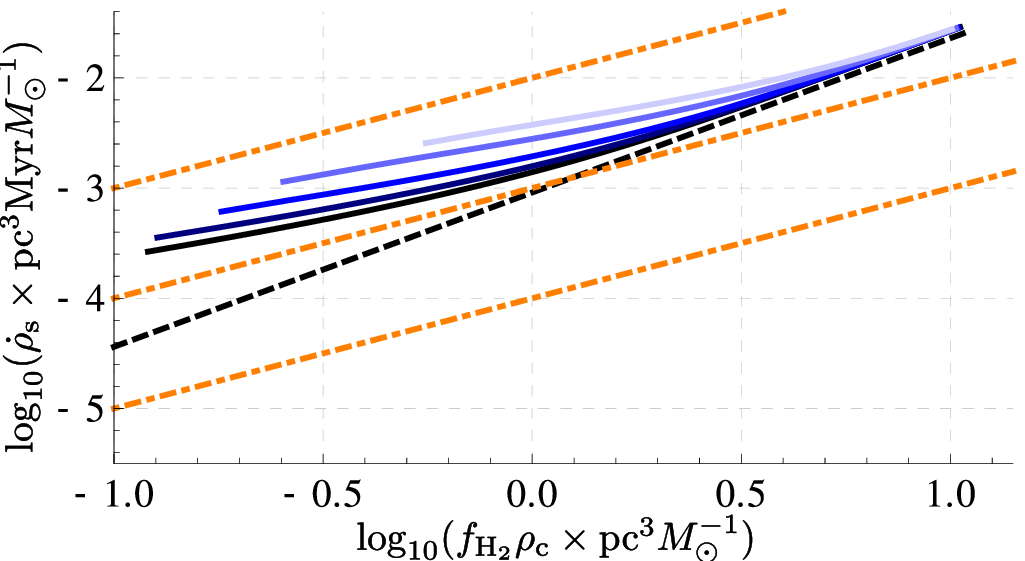}\hspace{18mm}
\includegraphics[width=70mm]{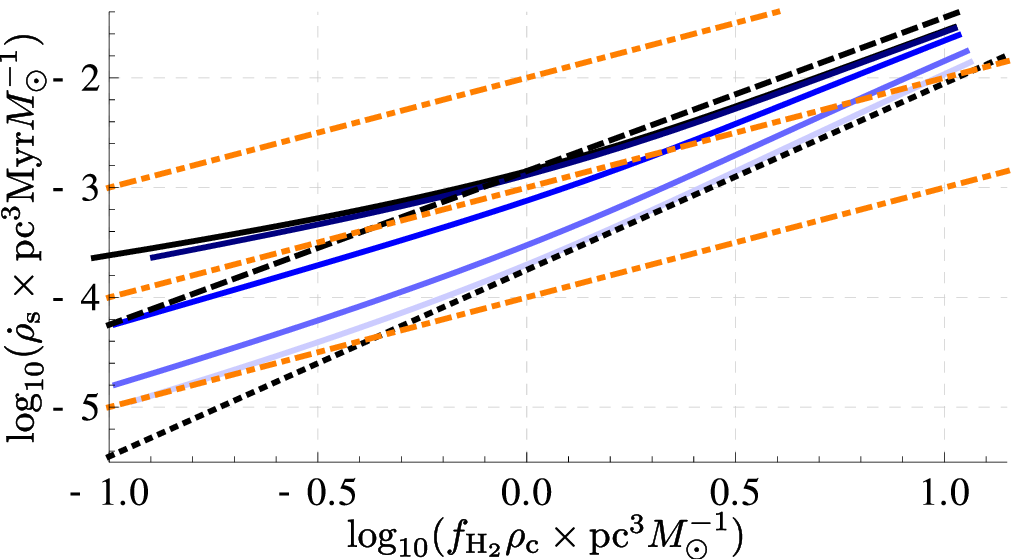}
\caption{Star formation rate as function of the fractional molecular density $f_{\rm c,H_2}\rho_{\rm c}$ for $Z/Z_\odot= 1.0,\,0.8,\,0.6,\,0.4,\,0.3$ from dark to light blue (left panel) and for different turbulent energy injection rates $\Sigma$ in terms of $\mathcal{M}_{\rm \Sigma}=0.0,\,0.4,\,0.8,\,1.2,\,1.4$ (from dark to light blue, right panel). The long- and short-dashed lines indicate the asymptotes with slopes $1.4$ ($\mathcal{M}_{\rm \Sigma}=0.0$, all $Z/Z_\odot$) and $1.6$ ($\mathcal{M}_{\rm \Sigma}=1.0$), respectively. Orange dot-dashed lines are as in Fig.~\ref{fig:rhoSdotVSrhoHI}.}
\label{fig:rhoSdotVSrhoH2}
\end{figure*}

Observed star formation surface densities and depletion timescales, as well as the related surface densities of molecular and atomic hydrogen, are projected quantities. In the simplest case of a face-on galactic disk, these quantities are averaged over the thickness of the disk and over a certain solid angle (or area). Since we can not account for spatial structures in the the equilibrium one-zone models, we plot of the star formation density over the atomic, molecular, and total volume densities (with helium and metals included) for different metallicities and varying external driving in  Figs.~\ref{fig:rhoSdotVSrhoHI}, \ref{fig:rhoSdotVSrhoH2}, and \ref{fig:rhoSdotVSrho}, respectively. However, these plots should reproduce observed trends because the typical thickness of a star forming region is of the order of $\gtrsim\,10\,\mathrm{pc}$, which is comparable to the typical length scale $\ell$ of our models, and the gas in star forming regions contribute most to column densities. Indeed, in comparison to observational rates and densities \citep[e.g.][Fig. 4 and Fig. 11,respectively]{Bigiel2008,Schruba2011}, we find a shift in numbers by a factor $\gtrsim\,10$ \footnote{The densitiy plots are in units of $M_\odot\mathrm{pc}^{-3}$ and $M_\odot\mathrm{Myr}^{-1}\mathrm{pc}^{-3}$, while the observational column density plot are usually in units of $M_\odot\mathrm{pc}^{-2}$ and $M_\odot\mathrm{yr}^{-1}\mathrm{kpc}^{-3}$.}, in both directions, but the general behaviour is very similar.

Figure~\ref{fig:rhoSdotVSrhoHI} shows that the star formation rate is clearly not correlated to $\mathrm{HI}$-densities below $\lesssim1\,M_\odot\mathrm{pc}^{-3}$ (roughly corresponding to column densites  $\lesssim10\,M_\odot\mathrm{pc}^{-2}$), and higher densities averaged over the cold and warm phases are atypical. The star formation rate as a function of the total gas density (see Fig.~\ref{fig:rhoSdotVSrho}) switches from zero at low densities, for which the fraction of $\mathrm{H_2}$ is neglibile, to a tight correlation above $\sim1\,M_\odot\mathrm{pc}^{-3}$. This threshold is caused by the transition from atomic to shielded molecular gas. Correspondingly, the star formation rate mainly correlates with the $\mathrm{H_2}$-density, which agrees with the KMT09 model. For a given set of parameters, the molecular gas depletion timescale $t_{\rm dep,H_2}=\dot{\rho}_{\rm s}/(f_{\rm c,H_2}\rho_{\rm c})$ varies only little over two orders of magnitude in $\mathrm{H_2}$-density (see Fig.~\ref{fig:rhoSdotVSrhoH2}). While the model predicts a depletion time of $\sim0.7\,\mathrm{Gyr}$ if internal driving is the dominant mechanism of turbulence production, the depletion time scale increases significantly for external turbulence driving. For example, \citet{Bigiel2011} find $\sim2\,\mathrm{Gyr}$ observationally, which could be maintained by external driving for $\mathrm{H_2}$-densities up to a few $M_\odot\mathrm{pc}^{-3}$. However, as we underestimate the $\mathrm{H_2}$-content in the gas, the depletion time tends to be too low in our model toward lower densities so that less external energy injection might be required to obtain the observed depletion time. A further caveat is the assumption of equilibrium. 

Eventually, only the application as an SGS model in numerical simulations of galaxies will enable us to calculate relations between the star formation rate and the surface density by projecting the computed density fields. Since star forming regions go through different evolutionary stages, during which star formation occurs episodically,  the star formation rate has to be integrated in time to incorporate non-equilibrium effects into the depletion time scale.

\section{Discussion and Conclusions}

In this paper we propose a model for the multi-phase ISM and star formation, considering the effects of turbulence and stellar feedback. Based on the concept of \citet{SpringHern03}, we split the gas content of a region of given size into a two distinct fractions, representing a diffuse warm and a clumpy cold component. However, our model goes significantly beyond their approach. By applying a simplified treatment of molecular hydrogen formation and destruction, we relate the star formation rate to the fractional density of molecular hydrogen in the cold-gas phase \citep{KrumKee09} [KMT09]. While star formation models that are applied in numerical simulations usually assume a constant efficiency parameter that is globally calibrated against the Kennicutt-Schmidt law, we dynamically calculate the star formation efficiency on the basis of local physical processes. In the spirit of \citet{KrumKee05}, star formation is regulated by the virial parameter and the turbulent velocity dispersion of cold clumps. The interrelationship in our model is more elaborate though. Turbulent energy can be produced by a turbulent cascade from larger scale, but also via internal driving by the thermal instability of the gas and by supernova feedback. To determine the gas fraction that collapses per free-fall timescale into stars, we assume a log-normal distribution of density fluctuations in the cold gas and we relate the critical density for a gravitational collapse to the virial parameter and the turbulent Mach number on the typical length scale of the cold clumps \citep{PadNord09} [PN11]. To account for the effect of SNe, we use a delayed feedback model. Apart from the turbulent energy, the fractional densities of the cold and warm phases and the thermal energy of warm gas are evolved (the temperature of the cold gas is assumed to be constant). Mass and energy is exchanged between the phases via radiative cooling, heating and mixing processes. An important source of heating is the stellar population in the volume. We consider two feedback mechanisms, taking the time scales of stellar evolution into account: the Lyman-continuum emission and SNII-explosions of young massive stars. To close the system of equations, we assume an effective (i.~e., thermal plus turbulent) pressure balance between the cold clumps and the surrounding warm gas.

By integrating the evolutionary equations for the averaged mass fractions and energies in a given spatial volume, we have obtained semi-analytical one-zone models, with the total gas density $n$, the metallicity $Z$ and the rate of energy injection by external turbulence forcing, $\Sigma$, as main parameters. Of particular interest are equilibrium solutions with a constant star formation rate. Fig.~\ref{fig:rhoSdotVSrho} shows that, we obtain asymptotic Kennicutt-Schmidt-relations with slope $1.5$ ($\dot{\rho}_{\rm s}=\rho\varepsilon_{\rm ff}/t_{\rm ff}\propto\rho^{1.5}$) toward high densities, which is a consequence of the asymptotically constant star formation efficiency $\varepsilon_{\rm ff}$. Depending on the metallicity and other parameters, the threshold densities are typically between $20$ and about $200\,\mathrm{cm^{-3}}$. For reasonable choices of the model coefficients that control internal turbulence driving and heating, a star formation efficiency of around 0.5\% is obtained above the threshold densities, in agreement with observed values \citep[e.g.][]{KrumTan07,Bigiel2008,Bigiel2011,Schruba2011,Onodera2010}. External turbulence driving (i.~e., energy transfer from larger scales via the turbulent cascade) decreases the star formation rate and slightly changes the slope of the power-law branches (Fig.~\ref{fig:rhoSdotVSrho}, right panel). This is primarily caused by the effect of the turbulent pressure on the average density of the cold phase, while the direct influence of turbulence on the star formation efficiency, following the prescription of PN11, plays a role in violently turbulent environments. In the latter case, also the production of molecular hydrogen fraction is affected via the turbulent clumping factor. Remarkably, the star formation efficiency is quite sensitive on the factor $b$ in Eq.~(\ref{eq:sigmapdf}) for the width of the density pdf as a function of the turbulent Mach number in the cold-gas phase. As shown by \citet{Federrath2010b}, $b$ is related to the mixture of the solenoidal and compressive components of the turbulent velocity field. Moreover, they concluded from comparisons with observed two-point statistics of turbulence in molecular clouds that this mixture varies for different clouds. Thus, it appears to be important to account for variations in the turbulence statistics.

Recent observations indicate a particularly tight correlation of the star formation rate with the molecular gas column densities in galaxies down to kpc scales (KMT09 give an overview of observational results). Since we consider local regions of the ISM with a size smaller than the galactic disc thickness, it is not reasonable to express the results from our one-zone models in terms of column densities. For the same reason, comparisons with the model of \citet{OstKee10} are difficult. Nevertheless, we find that the relation between the equilibrium star formation rate and the density of molecular hydrogen closely follows a power law, particularly for solar metallicity. As one can see in the left panel of Fig.~\ref{fig:rhoSdotVSrhoH2}), the star formation rate is $\dot{\rho}_{\rm s}\propto\rho^{1.5}$ for sufficiently high density. Strong external turbulence forcing significantly reduces the star formation rate and the slope of the asymptote increases from $1.4$ to about $1.6$ (Fig.~\ref{fig:rhoSdotVSrhoH2}, right panel). In this regard, it is interesting that KMT09 distinguish two different regimes, in which molecular clouds are either self-regulated (at low surface densities) or significantly affected by their galactic environment (at high surface densities). In the former case, they derive a slope of about $1.4$, whereas the slope is about $1.6$ in the latter case. A plausible interpretation in the context of our model is that these regimes roughly correspond to internal turbulence driving as the dominating production mechanism (negligible $\Sigma$) vs.\ significant turbulence production by the transport from instabilities on large scale to molecular cloud scales (large $\Sigma$). To corroborate this interpretation, the model has to be applied in simulations of disc galaxies. The modelled equilibrium star formation rates and depletion time scales are roughly consistent with those found observationally \citep[e.g.][]{Schruba2011}, and the modelled relation between star formation rate and molecular gas density is in agreement with a more or less constant molecular gas depletion time scale as observed by \citet{Bigiel2011}.

In such simulations as well as in cosmological simulations, it is common to assume a constant star formation efficiency beyond a certain threshold density. The results of our numerical study suggest that this is a reasonable approximation. However, rather than using this as an entirely phenomenological input to the simulations, the equilibrium values of the star formation efficiency calculated with our model follow from the sub-resolution physics of the ISM. Moreover, rather than setting stiff density thresholds, the model yields transition values depending on the varying gas density in numerical simulations. To utilise the equilibrium solutions as a parametrization of star formation, tables of the star formation efficiency as function of density and metallicity can be calculated (the code calculating the efficiencies can be an be obtained from the authors upon request). The rate of external turbulent energy production could be estimated, for instance, from the large-scale velocity dispersion in galaxies \citep[see][]{BurkGenz09}. 

While such a simplified approach has its merits, it cannot account for dynamical effects. A crucial problem is the calculation of the local rate of turbulent energy production on the grid scale due the shear of numerically resolved turbulent flow in a simulation (i.~e., the energy transfer from length scale greater than the size of the grid cells to unresolved length scales). This is the meaning of $\Sigma$ if the proposed model is applied as a sub-grid scale model. Since the turbulent velocity fluctuations in the ISM can assume a significant fraction of the sound speed or even become supersonic, an incompressible turbulence model is not sufficient. \citet{SchmFed10} [SF11] provide a formula for $\Sigma$ in the highly compressible regime. A complete model for the turbulent multi-phase ISM and star formation is obtained by rewriting Eqs. ~(\ref{eq:raterhoS}, \ref{eq:raterhoC}, \ref{eq:raterhoH}, \ref{eq:rateuH}, \ref{eq:rateeSGS}, \ref{eq:rateZ}) as partial differential equations with fluid-dynamical advection terms, where the length scale $l$ is given by size of the grid cells, $\Delta$, and $e_{\rm t}=e_{\rm sgs}$ is identified with the unresolved fraction of the kinetic energy (see SF11). These equations supplement the Euler equations for the total gas density, the momentum, and the total energy. Solving the complete set of equations will be a substantial numerical challenge.

To perform simulations of galaxies in cosmological environments, adaptive mesh refinement (AMR) is indispensable. \citet{MaierIap09} incorporated an SGS turbulent energy equation for moderately compressible turbulence into AMR simulations of galaxy clusters. This method also can be applied using the multi-phase model for the turbulent ISM. Then the length scale $l$ of the model corresponds to the varying grid scale, and the scale-dependent turbulent energy has to be adjusted if refined grids are inserted or solutions on finer grids are projected to coarser grid levels.

An advantage of our model is that components can be modified, replaced and added as our understanding of the physics of the ISM progresses. For example, the computation of the dimensionless star formation rate from the turbulent cold-gas density pdf (see Sect.~\ref{sec:SFR}) is more or less heuristic. We anticipate that theoretical advances and results form small-scale simulations will soon lead to improvements. An important issue we have not considered so far is the influence of magnetic fields. The role of MHD turbulence is already emphasized by PN11. Although additional complications arise when magneto-turbulent fluctuations have to be treated on sub-grid scales, it is a problem that can be tackled. Furthermore, a more detailed treatment of chemical processes is desirable, although \citet{KrumGned11} have demonstrated that the simple analytical model for the molecular hydrogen fraction in KMT09 agrees quite well with an explicit reaction network in cosmological simulations, at least if the metallicity is not much lower than solar. The multi-phase model also has to be adapted to the simulation framework. For cosmological simulations with relatively coarse resolutions, the neutral gas phases should be embedded in a hot ionised medium of low density. The other extreme are simulations of isolated disc galaxies with very high resolution, in which cold clumps can be marginally resolved so that several neighbouring grid cells are completely filled by cold gas. In this case, it is necessary to switch from the two-phase description to the one-phase limit.

The predictive power of astrophysical simulations, in which the ISM is only partially resolved, will increase by applying the equilibrium solutions or by implementing the full multi-phase SGS model. This will allow us, in turn, to test and to modify the underlying physical assumptions. 

\section*{Acknowledgments}

Harald Braun was financially supported by the DFG project Ni 516/701. Wolfram Schmidt is grateful for discussions with the participants of the 2010 summer program \textit{Star Formation in Galaxies: From Recipes to Real Physics} at the Aspen Center for Physics, which helped to bring this work to maturity. We thank Jens Niemeyer for supporting this project and for many comments that helped to improve this paper. We also thank the referee for many helpful suggestions and Christoph Federrath for further comments. 
\bibliography{TUandSFModBib}

\appendix 
\bsp

\label{lastpage}

\end{document}